\documentclass[11pt]{article}
\usepackage{amsfonts}
\usepackage{mathrsfs}
\usepackage{graphicx}
\usepackage{amsmath}
\usepackage{amssymb}
\usepackage{subfigure}
\usepackage{epsfig}
\usepackage{graphicx}
\usepackage{color}
\usepackage{indentfirst}
\usepackage{hyperref}
\usepackage{setspace}
\usepackage{braket}
\usepackage{slashed}
\usepackage{cite}
\usepackage{multirow}
\usepackage{cancel}
\usepackage{caption}
\usepackage{array}
\usepackage{appendix}
\usepackage{tabularx}
\usepackage{tabulary}

\hypersetup{colorlinks=true, citecolor=red, urlcolor=blue, linkcolor=blue}
\begin{document}
\title{Rare $ \Lambda_c $ decays and new physics effects}
\author{Sheng-Qi Zhang$ ^{1} $, and Cong-Feng Qiao$^{1}\footnote{qiaocf@ucas.ac.cn.}$\\
\\
\normalsize{$^{1}$School of Physical Sciences, University of Chinese Academy of Sciences}\\
\normalsize{YuQuan Road 19A, Beijing 100049, China}\\
\\
}
\date{}
\maketitle

\begin{abstract}
Recent experimental progress on baryonic rare decays has spurred a deeper investigation on flavor-changing neutral current transitions in the baryon sector. Within the framework of QCD sum rules, we derive a complete set of form factors for the $ \Lambda_c\to p $ process in the large recoil region and use the $z$-series parametrization to extrapolate them across the full physical range. Employing these form factors and flavor symmetries, we compute branching fractions for the decays $\Lambda_c \to p e^+ e^-$ and $\Lambda_c \to p \mu^+ \mu^-$, as well as for rare $ \Xi_c $ decay modes. We examine as well the new physics effects through specific angular observables such as the lepton forward-backward asymmetry and the fraction of longitudinally polarized dileptons. Results indicate that new physics models may be testified in baryonic rare decays, with immense data collected in running and future colliders.
\vspace{0.3cm}
\end{abstract}

\section{INTRODUCTION}

Rare flavor-changing neutral current (FCNC) decays, induced by $ b \to s \ell \ell $, $ s \to d \ell \ell $, and $ c \to u \ell \ell $, are excellent candidates for probing new physics due to their severe suppression by the Glashow-Iliopoulos-Maiani mechanism within the Standard Model (SM)~\cite{Glashow:1970gm}. These decays are prohibited at the tree level and occur only through loop diagrams. Flavor-changing neutral current processes enable precise determination of the Cabibbo-Kobayashi-Maskawa (CKM) matrix elements, which can be used to rigorously test the SM and its possible extensions. Additionally, because their amplitudes can interfere with contributions from non-SM particles, these rare decays provide a unique opportunity to explore the existence of yet unobserved particles, including supersymmetry and dark matter candidates~\cite{Hewett:1996ct, Buchalla:2000sk, Bird:2004ts, Davidson:1993qk}.

Experimental analysis of the rare decays are predominantly focused on the meson system~\cite{Belle:2001oey, BaBar:2006tnv, BaBar:2008jdv, Belle:2009zue, LHCb:2012juf, LHCb:2014cxe, BELLE:2019xld, LHCb:2019hip, BaBar:2011ouc, LHCb:2013hxr, LHCb:2020car, CLEO:2010ksb, D0:2007qbl, NA482:2009pfe, NA62:2022qes}, while baryonic rare decays remain less explored due to the complexity of baryon structures and their comparatively lower production rates. However, baryonic decays have distinct advantages. They allow the extraction of helicity structures from the effective Hamiltonian, which are often obscured in mesonic decays~\cite{Mannel:1997xy}. Furthermore, baryonic decays involve a diquark system as a spectator, unlike mesonic decays where a single quark is involved, leading to different hadronic physics. Baryonic decays may also exhibit contributions from $ W $ exchange involving both two-quark and three-quark transitions~\cite{Wang:2021uzi}. Hence, baryonic rare decays provide crucial complementary information for the mesonic processes. 

To date, only a few baryonic rare decay channels involving $ b\to s\ell\ell $ and $ s\to d\ell\ell $ have been observed experimentally~\cite{CDF:2011buy, LHCb:2013uqx, LHCb:2017rdd, NA48:2007smd}. In contrast, up-type quark transitions $ c\to u\ell\ell $ remain unobserved. Recently, the LHCb collaboration set an upper limit on the branching fraction of $ \Lambda_c \to p \mu^+ \mu^- $\cite{LHCb:2024hju} to be $3.2 \times 10^{-8}$, improving upon previous results from the \textit{BABAR} and LHCb\cite{BaBar:2011ouc, LHCb:2017yqf}. Transitions involving $c \to u\ell \ell $ are subject to significantly stronger Glashow-Iliopoulos-Maiani suppression compared to the bottom counterparts, leading to extremely small branching fractions\cite{deBoer:2015boa, Fajfer:2001sa, Paul:2011ar}. Consequently, rare charm decays are relatively uncharted territory both theoretically and experimentally, which makes such studies desirable.

Probing new physics in rare decays can be achieved by measuring angular observables that are particularly sensitive to non-SM effects. For instance, the forward-backward asymmetry in $ \Lambda_c\to p\ell\ell $ exhibit an heavily dependence on the value of Wilson coefficient $ C_{10} $, which is predicted to be zero within the SM~\cite{Golz:2021imq}. Moreover, deviations from the SM predictions of the angular observables have been already observed in the $ b\to s\ell\ell $ transitions~\cite{LHCb:2020gog}, making it essential to examine similar effects in the $ c\to u\ell\ell $ case. However, before attributing any observed anomalies to new physics, careful scrutiny of experimental data and SM predictions is required to ensure that these deviations are not due to overlooked factors within the SM itself, such as higher-order corrections or soft-gluon effects~\cite{Wang:2008sm}.

Theoretically, description of baryonic rare decays relies on the effective Hamiltonian, where intermediate gauge bosons are integrated out. Using the operator product expansion (OPE), the dilepton and hadronic parts can be separated. The dilepton component is governed by the Wilson coefficients, whereas the hadronic component is generally parametrized in terms of form factors. At present, the form factors for baryonic rare decays are mainly derived from several phenomenological models, such as the MIT bag model~\cite{Cheng:1994kp}, the relativistic quark model (RQM)~\cite{Faustov:2017wbh, Faustov:2017ous, Faustov:2018dkn}, the covariant constituent quark model~\cite{Gutsche:2013pp}, the Bethe-Salpeter equation approach~\cite{Liu:2019igt, Liu:2019rpm}, the light-cone sum rules (LCSR)~\cite{Wang:2008sm, Azizi:2010zzb, Gan:2012tt, Aliev:2018hyy}, and the lattice QCD (LQCD)~\cite{Detmold:2016pkz, Meinel:2017ggx}. Additionally, the heavy quark effective theory (HQET) is commonly applied to simplify form factor calculations, particularly in the context of the $b$-quark systems~\cite{Hussain:1992rb, Mannel:1990vg, Chen:2001ki, Huang:1998ek}. For example, Huang \textit{et al}. utilized the QCD sum rules method to determine two HQET form factors of $ \Lambda_b \to \Lambda \ell \ell $ decay mode and computed the corresponding decay width~\cite{Huang:1998ek}. The QCD sum rules (QCDSR) method they used, has been extensively applied to study the decay properties of hadrons in the past few decades, including both strong~\cite{Lian:2023cgs, Wang:2016wkj, Dias:2013qga, Wan:2020oxt, Wang:2023sii} and semileptonic decays~\cite{Dai:1996xv, Zhang:2023nxl, Huang:1998rq, Huang:1998ek, MarquesdeCarvalho:1999bqs, Shi:2019hbf, Zhao:2020mod, Zhao:2021sje, Xing:2021enr, zhang:2024ick}. Unlike phenomenological models, QCDSR is a theoretical framework grounded in the first principles of QCD. This method begins by constructing three-point correlation functions using suitable interpolating current operators, with nonperturbative effects captured as vacuum expectation values across various dimensions. Once the equivalence between the QCD and phenomenological representations of the three-point correlation functions is established through quark-hadron duality, the form factors associated with weak transitions are rigorously determined. In this work, we employ QCDSR to calculate the complete set of the form factors of the rare decay mode $ \Lambda_c\to p\ell\ell $, after which the branching fractions as well as the angular observables are obtained. The potential new physics effects are also analyzed. Besides, by exploiting certain flavor symmetry, rare $ \Xi_c $ decay modes can be investigated.

The paper is organized as follows: in Sec.~\ref{sec:Formalism} we  explain the foundational concept of QCDSR for three-point correlation functions. The numerical results along with a detailed analysis are presented in Sec.~\ref{sec:Numerical}. Finally, the conclusions and discussions are provided in the last section.

\section{FORMALISM}\label{sec:Formalism}

The amplitudes of $ \Lambda_c\to p\ell\ell $ are derived by calculating the hadronic matrix element between the $ \Lambda_c $ and proton ($ p $). In general, the hadronic matrix element can be parametrized in terms of the form factors\cite{Faustov:2017wbh, Xing:2018lre}:
\begin{align}
\bra{p(q_2)}\bar{u}\gamma_\mu\left(1-\gamma_5\right)c\ket{\Lambda_c(q_1)}&=\bar{u}_{2}(q_2) \bigg[f_1(q^2) \gamma_\mu+i f_2(q^2)\frac{\sigma_{\mu\nu}\,q^\nu}{M_{\Lambda_c}}+f_3(q^2) \frac{q_\mu}{M_{\Lambda_c}} \bigg] u_1(q_1)\nonumber\\
&-\bar{u}_{2}(q_2)\bigg[g_1(q^2) \gamma_\mu+i g_2(q^2) \frac{\sigma_{\mu\nu}\,q^\nu}{M_{\Lambda_c}}+g_3(q^2) \frac{q_\mu}{M_{\Lambda_c}}\bigg]\gamma_5 u_1(q_1)\;,\nonumber\\
\bra{p(q_2)}\bar{u}i\sigma_{\mu\nu}q^\nu\left(1+\gamma_5\right)c\ket{\Lambda_c(q_1)}&=\bar{u}_{2}(q_2) \bigg[\frac{f_1^{T}(q^2)}{M_{\Lambda_c}}\left(\gamma^\mu q^2-q^\mu\slashed{q}\right)-f_2^{T}(q^2)i\sigma_{\mu\nu}q^\nu  \bigg] u_1(q_1)\nonumber\\
&+\bar{u}_{2}(q_2) \bigg[\frac{g_1^{T}(q^2)}{M_{\Lambda_c}}\left(\gamma^\mu q^2-q^\mu\slashed{q}\right)-g_2^{T}(q^2)i\sigma_{\mu\nu}q^\nu  \bigg]\gamma_5 u_1(q_1)\;,
\label{matrix}
\end{align}
where $ q_{1} $ and $ q_2 $ denote the the four-vector momentum of the initial and final state baryons and the momentum transfer is defined as $ q = q_1-q_2 $. To calculate the above form factors in the framework of QCDSR, the three-point correlation functions could be constructed as
\begin{equation}
\label{3pcf}
\Pi_\mu(q_1^2, q_2^2, q^2)=i^2 \int d^4 x\; d^4 y\; e^{i(-q_1 x+q_2 y)} \bra{0}T\{j_{p}(y)j^{I,II}_\mu(0)j^{\dagger}_{\Lambda_c}(x) \}\ket{0}\;.
\end{equation}

Here, two weak transition currents and the hadron currents take the following quark structure~\cite{Chung:1981wm, Zhao:2020mod}:
\begin{align}
\label{current-weak1}
j_\mu^I &= \bar{u}\gamma_\mu\left(1-\gamma_5\right)c\;,\\
\label{current-weak2}
j_\mu^{II}&=\bar{u}i\sigma_{\mu\nu}q^\nu\left(1+\gamma_5\right)c\;,\\
\label{current-Lambdac}
j_{\Lambda_c}&=\epsilon_{ijk}(u_i^TC\gamma_5d_j)c_k\;,\\
\label{current-p}
j_p&=\epsilon_{ijk}(u_i^TC\gamma_5d_j)u_k\;,
\end{align}
where the subscripts $i$, $j$, and $k$ denote the color indices, and $C$ represents the charge conjugation matrix.

On the phenomenological side, by inserting a complete set of intermediate hadronic states and applying double dispersion relations, the three-point correlation functions in Eq.~(\ref{3pcf}) can be expressed as
\begin{align}
\label{spectra}
\Pi_\mu^{\text{phe}}(q_1^2, q_2^2, q^2)=&\sum_{\text{spins}}\frac{\bra{0}j_{p}\ket{p(q_2)}\bra{p(q_2)}j^{I,II}_\mu\ket{\Lambda_c(q_1)}\bra{\Lambda_c(q_1)}j_{\Lambda_c}\ket{0}}{(q_1^2-M_{\Lambda_c}^2)(q_2^2-M_{p}^2)}\nonumber\\
+&\text{higher resonances and continuum states}\;.
\end{align}

Here, the transition amplitudes from the vacuum to baryon states can be parametrized by introducing the decay constants:
\begin{align}
\bra{0}j_{\Lambda_c}\ket{\Lambda_c(q_1)}&=\lambda_{\Lambda_c}u_{\Lambda_c}(q_1)\;,\\
\bra{0}j_{p}\ket{p(q_2)}&=\lambda_{p}u_{p}(q_2)\;,
\end{align}
where $ \lambda_{\Lambda_c} $ and $ \lambda_{p} $ represent the decay constants of $ \Lambda_c $ and the proton, respectively. Introducing the hadronic transition matrix elements in Eq.~(\ref{matrix}) and utilizing the spin sum completeness relations, $ \sum u_{\Lambda_c}(q_1)\bar{u}_{\Lambda_c}(q_1)=\slashed{q}_1+M_{\Lambda_c} $ and $ \sum u_{p}(q_2)\bar{u}_{p}(q_2)=\slashed{q}_2+M_{p} $, one can derive the phenomenological representation of the three-point correlation functions in Eq.~(\ref{3pcf}),
\begin{align}
\label{3ptphe1}
\Pi_\mu^{I,\,\text{phe}}(q_1^2, q_2^2, q^2) & =\frac{\lambda_{p} \left(\slashed{q}_2+M_{p}\right)\bigg[f_1(q^2) \gamma_\mu+i f_2(q^2)\frac{\sigma_{\mu\nu}\,q^\nu}{M_{\Lambda_c}}+f_3(q^2)\frac{q_\mu}{M_{\Lambda_c}}\bigg]\lambda_{\Lambda_c}\left(\slashed{q}_1+M_{\Lambda_c}\right)}{(q_1^2-M_{\Lambda_c}^2)(q_2^2-M_{p}^2)} \nonumber\\
&-\frac{\lambda_{p} \left(\slashed{q}_2+M_{p}\right)\bigg[g_1(q^2) \gamma_\mu+i g_2(q^2)\frac{\sigma_{\mu\nu}\,q^\nu}{M_{\Lambda_c}}+g_3(q^2)\frac{q_\mu}{M_{\Lambda_c}} \bigg]\gamma_5\lambda_{\Lambda_c}\left(\slashed{q}_1+M_{\Lambda_c}\right)}{(q_1^2-M_{\Lambda_c}^2)(q_2^2-M_{p}^2)}\nonumber\\
&+\text{higher resonances and continuum states}\;,
\end{align}
\begin{align}
\label{3ptphe2}
\Pi_\mu^{II,\,\text{phe}}(q_1^2, q_2^2, q^2) & =\frac{\lambda_{p} \left(\slashed{q}_2+M_{p}\right)\bigg[\frac{f_1^{T}(q^2)}{M_{\Lambda_c}}\left(\gamma^\mu q^2-q^\mu\slashed{q}\right)-f_2^{T}(q^2)i\sigma_{\mu\nu}q^\nu\bigg]\lambda_{\Lambda_c}\left(\slashed{q}_1+M_{\Lambda_c}\right)}{(q_1^2-M_{\Lambda_c}^2)(q_2^2-M_{p}^2)} \nonumber\\
&+\frac{\lambda_{p} \left(\slashed{q}_2+M_{p}\right)\bigg[\frac{g_1^{T}(q^2)}{M_{\Lambda_c}}\left(\gamma^\mu q^2-q^\mu\slashed{q}\right)-g_2^{T}(q^2)i\sigma_{\mu\nu}q^\nu \bigg]\gamma_5\lambda_{\Lambda_c}\left(\slashed{q}_1+M_{\Lambda_c}\right)}{(q_1^2-M_{\Lambda_c}^2)(q_2^2-M_{p}^2)}\nonumber\\
&+\text{higher resonances and continuum states}\;.
\end{align}

On the QCD side, the three-point correlation functions shown in Eq.~(\ref{3pcf}) can be expressed using the OPE and double dispersion relations:
\begin{align}
\label{3ptQCD}
\Pi_\mu^{\text{QCD}}(q_1^2, q_2^2, q^2)=\int_{s_1^\text{min}}^{\infty}d s_1\int_{s_2^\text{min}}^{\infty}d s_2\frac{\rho^{\text{QCD}}_\mu(s_1,s_2,q^2)}{(s_1-q_1^2)(s_2-q_2^2)}\;,
\end{align}
where $ s_{1(2)}^\text{min} $ denotes the kinematic limit. The spectral density $ \rho^{\text{QCD}}_\mu(s_1,s_2,q^2) $ can be derived by employing Cutkosky cutting rules~\cite{Zhao:2020mod, MarquesdeCarvalho:1999bqs,Zhao:2021sje, Shi:2019hbf, Xing:2021enr, Wang:2012hu, Yang:2005bv, Du:2003ja}. In this work, contributions up to dimension six are taken into account for $ \rho^{\text{QCD}}_\mu(s_1,s_2,q^2) $ and expressed as:
\begin{align}
\label{spectra-density}
\rho^{\text{QCD}}_\mu(s_1,s_2,q^2)&=\rho^{\text{pert}}_\mu(s_1,s_2,q^2)+\rho^{\langle \bar{q}q\rangle}_\mu(s_1,s_2,q^2)+\rho^{\langle g_s^2G^2 \rangle}_\mu(s_1,s_2,q^2)\nonumber\\
&+\rho^{\langle g_s \bar{q}\sigma \cdot  G q \rangle}_\mu(s_1,s_2,q^2)+\rho^{\langle \bar{q}q \rangle^2}_\mu(s_1,s_2,q^2)\;.
\end{align}
Here the first term represents to the perturbative contribution, whereas $ \langle\bar{q}q \rangle$, $  \langle g_s^2G^2 \rangle $, $ \langle g_s \bar{q}\sigma \cdot  G q \rangle $, and $ \langle \bar{q}q \rangle^2 $ reflect nonperturbative effects, each corresponding to a different order of vacuum expectation values known as condensates.

To establish the equivalence between phenomenological and QCD representations, one should employ quark-hadron duality and the double Borel transform. As a result, the general form of the form factors can be derived:
\begin{align}
\label{form-factors-exp}
f^{\Lambda_c\rightarrow p}(q^2)=\frac{e^{M_{\Lambda_c}^2/\tau_{1}^2}e^{M_p^2/\tau_{2}^2}}{\lambda_{\Lambda_c}\lambda_p}\int_{s_1^\text{min}}^{s_1^0}d s_1\int_{s_2^\text{min}}^{s_2^0}d s_2\;\rho(s_1,s_2,q^2)e^{-s_1/\tau_1^2} e^{-s_2/\tau_2^2}+\Pi_{0}\;.
\end{align}
Here, $ s_1^0 $ and $ s_2^0 $ represent the threshold parameters of the $ \Lambda_c $ and proton, respectively. $ \Pi_0 $ includes contributions to the correlation function that lack an imaginary part but exhibit nontrivial features after the Borel transform. Additionally, $ \tau_{1}^2 $ and $ \tau_{2}^2 $ denote the Borel parameters that appear following the double Borel transform.

\section{NUMERICAL RESULTS
AND DISCUSSIONS}\label{sec:Numerical}
\subsection{Calculation of form factors}
To calculate the form factors numerically, we adopt the following input parameters~\cite{Shifman:1978by, Wan:2022xkx, Colangelo:2000dp, Du:2003ja, Yang:2005bv, ParticleDataGroup:2024cfk, Wan:2021vny, Khodjamirian:2011jp, Chung:1984gr, Wan:2024fam, Wan:2023epq, Zhang:2022obn},
\begin{align}
\label{parameter}
\langle\bar{q}q\rangle &= -(0.24 \pm 0.01)^3 \text{GeV}^3\;, \nonumber \\
s_1^0 &= (9.5 \sim 10.5) \,\text{GeV}^2\;, \quad s_2^0 = (2.4\sim 3.0) \,\text{GeV}^2\;,\nonumber\\
m_c &= 1.27 \pm 0.02 \,\text{GeV}, \quad m_u = 2.16\pm 0.07 \,\text{MeV}\;, \nonumber \\
\lambda_{\Lambda_c} &= 0.0119 \,\text{GeV}^3\;, \quad \lambda_{p} = 0.02 \,\text{GeV}^3\;,\nonumber\\
M_{\Lambda_c} &= 2.286 \,\text{GeV}\;, \quad M_p = 0.938 \,\text{GeV}\;,
\end{align}	
where the decay constants and threshold parameters are obtained using the two-point sum rules~\cite{Khodjamirian:2011jp, Chung:1984gr}, employing the same interpolating currents as specified in Eqs.~(\ref{current-Lambdac}) and (\ref{current-p}). Additionally, two additional free parameters, $ \tau_1^2 $ and $ \tau_2^2 $, are introduced as described in Eq.~(\ref{form-factors-exp}). For simplicity, we utilize the following relation for the Borel parameters\cite{Shi:2019hbf, MarquesdeCarvalho:1999bqs, Leljak:2019fqa}:
\begin{align}
\frac{\tau_1^2}{\tau_2^2}=\frac{M_{\Lambda_c}^2-m_c^2}{M_{p}^2-m_u^2}\;.
\end{align}

Typically, two criteria are used to determine the values of the Borel parameters. The first criterion relates to the pole contribution. To ensure that the contribution of ground-state hadrons is significant, the pole contribution must dominate the spectrum. Consequently, a pole contribution greater than 40\% is often chosen for the transition form factors, as formulated below:
\begin{align}
R^{\text{PC}}_{\Lambda_c}=\frac{\int_{s_1^\text{min}}^{s_1^0}d s_1\int_{s_2^\text{min}}^{s_2^0}d s_2}{\int_{s_1^\text{min}}^{\infty}d s_1\int_{s_2^\text{min}}^{s_2^0}d s_2}\;,\\[5pt]
R^{\text{PC}}_{p}=\frac{\int_{s_1^\text{min}}^{s_1^0}d s_1\int_{s_2^\text{min}}^{s_2^0}d s_2}{\int_{s_1^\text{min}}^{s_1^0}d s_1\int_{s_2^\text{min}}^{\infty}d s_2}\;.
\end{align}
These two ratios can be regarded as the pole contribution from the $ \Lambda_c $ channel and proton channel, respectively.

The second criterion addresses the convergence of the OPE, which ensures the validity of the truncated OPE. In Eq.~(\ref{spectra-density}), nonperturbative effects up to dimension six are considered, meaning that the relative contribution from the highest dimension condensate $ \langle \bar{q}q\rangle^2 $ should be below 30\%. Additionally, as physical observables, form factors should ideally be independent of any artificial parameters. Therefore, a reliable result is obtained within an optimal region where the form factors show minimal dependence on $ \tau_{1}^2 $ and $ \tau_{2}^2 $. Practically, we adjust the threshold parameters $ s_{1,2}^0 $ by $ 0.1\, \text{GeV}^2 $ to determine the acceptable range of the Borel parameters.

With the above setup, the form factors for $ \Lambda_c \to p \ell\ell $ can be numerically determined. The results for the form factors at the maximum recoil point $ q^2 = 0 $, directly evaluated using QCD sum rules, are presented in Table~\ref{table:f0}, where the uncertainties arise from variations in the Borel parameters and the input parameters listed in Eq.~(\ref{parameter}). We also compared our results with predictions from other approaches and found that most results show reasonable agreement except $ f_3(0) $, $ f_1^T(0) $, and $ g_1^T(0) $. Since the current lack of experimental data and theoretical studies, further research on these form factors is necessary. Additionally, we observe that $ f_2^T(0) $ is equal to $ g_2^T(0) $, which aligns with the conclusion in Ref.~\cite{Gutsche:2013pp}.

\begin{table}[ht]
\centering
\caption{Theoretical predictions for the form factors of the rare decay $ \Lambda_c\to p \ell \ell $ at the maximum recoil point $ q^2=0 $, obtained using different approaches.}
\resizebox{\linewidth}{!}{
\begin{tabular}{lccccc}
\hline\hline
Method & $ f_1(0) $   & $ f_2(0) $   & $ f_3(0) $  & $f_1^T(0)$ & $f_2^T(0)$ \\ \hline
This work& $0.54 \pm 0.04$ & $-0.25\pm 0.02$ & $ 0.73\pm 0.09 $ & $ -0.62\pm 0.06 $  & $ -0.35\pm 0.04 $  \\
RQM\cite{Faustov:2018dkn}& $0.627$ & $-0.259$ & $ 0.179 $ & $ -0.310 $ & $ -0.380 $  \\
LCSR~\cite{Khodjamirian:2011jp} & $0.59^{+0.15}_{-0.16}$   & $-0.43^{+0.13}_{-0.12}$&  & &      \\
LQCD~\cite{Meinel:2017ggx} & $0.672\pm 0.039$   & $-0.321\pm 0.038$ &  & &\\     
\hline
& $ g_1(0) $ & $ g_2(0) $ & $g_3(0)$ & $g_1^T(0)$ & $g_2^T(0)$ \\ \hline
This work& $0.54 \pm 0.04$  & $-0.25\pm 0.02$ & $ -0.73\pm 0.09 $& $ 0.51\pm 0.12 $  & $ -0.35\pm 0.04 $  \\
RQM\cite{Faustov:2018dkn}& $0.627$  & $-0.259$ & $ -0.744 $ & $ 0.202 $ & $ -0.388 $  \\
LCSR~\cite{Khodjamirian:2011jp} & $0.55^{+0.14}_{-0.15}$   & $-0.16^{+0.08}_{-0.05}$& & &      \\
LQCD~\cite{Meinel:2017ggx} & $0.602\pm 0.031$  & $0.003 \pm 0.052$ &  & &\\     
\hline\hline 
\end{tabular}
}
\label{table:f0}
\end{table}

\begin{figure}[ht]
\centering
\includegraphics[width=7.9cm]{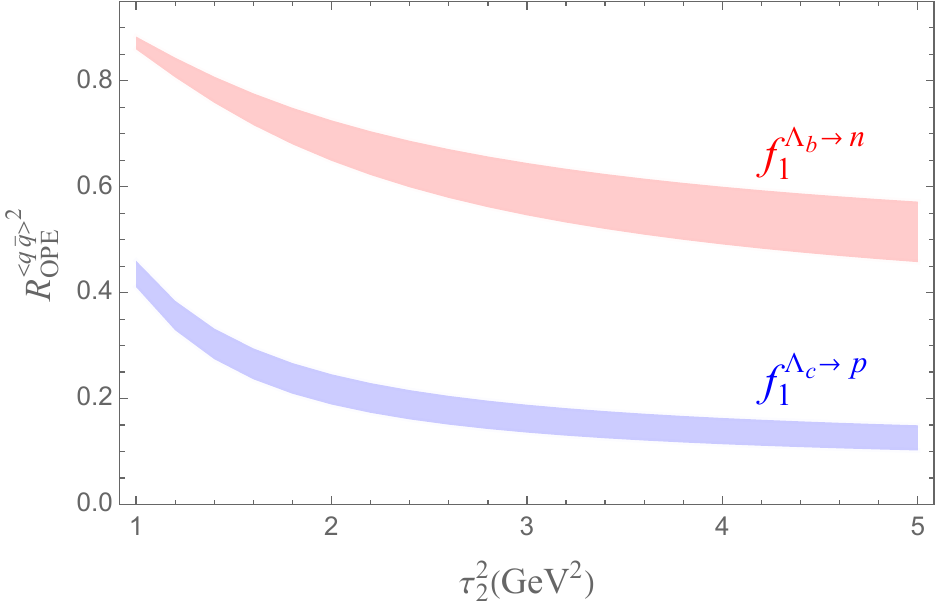}
\caption{The relative contributions of the highest dimension condensate $ \langle \bar{q}q\rangle^2 $ (red for $ \Lambda_b\to n\ell\ell $, blue for $ \Lambda_c\to p\ell\ell $) as functions of the Borel parameters $ \tau_2^2 $, where the error bands denote the uncertainties from threshold parameters.}
\label{fig:OPE}
\end{figure}

Notably, by substituting the $c$ quark in the $ \Lambda_c $ current~(\ref{current-Lambdac}) with a $b$ quark, the second $u$ quark in the proton current~(\ref{current-p}) with a $d$ quark, and adjusting the weak transition currents~(\ref{current-weak1}), (\ref{current-weak2}) as well as the hadronic matrix element~(\ref{matrix}) accordingly, the $ \Lambda_b\to n\ell\ell $ decay process can also be investigated. However, our calculations revealed that the highest-order condensate term $ \langle \bar{q}q\rangle^2 $ significantly contributes to the form factors. In Fig.~\ref{fig:OPE}, $R^{\langle \bar{q}q\rangle^2}_{\text{OPE}}$ illustrates the relative contribution of $ \langle \bar{q}q\rangle^2 $ to the form factors. It shows that within the permissible range of Borel parameters, the contribution of $ \langle \bar{q}q\rangle^2 $ can reach 50\%-90\% in $ \Lambda_b\to n\ell\ell $ decay process, which severely compromises the validity of the OPE. In contrast, the contribution of $ \langle \bar{q}q\rangle^2 $ in the $ \Lambda_c\to p\ell\ell $ process is much smaller. Our result is consistent with the conclusion in Ref.~\cite{Ball:1997rj}, which states that three-point QCDSR has some general limitations in the context of $ b $-quark decays. Specifically, the coefficient of the $ \langle \bar{q}q\rangle^2 $ condensate terms increases more rapidly with $ m_b $ than the coefficient of the perturbative contribution, an effect that does not manifest in the two-point QCDSR~\cite{Khodjamirian:1997lay}. Therefore, the QCDSR method appears unreliable for $b$-hadron decay due to the nonconvergence of the OPE. An alternative approach is to consider the expansion of the correlation function in the light cone, known as light cone sum rules. Additionally, QCDSR method can be applied within the framework of HQET to address $ b $-hadron decays, as demonstrated in Refs~\cite{Huang:1998ek, Huang:1998rq}.

\begin{table}[ht]
\centering
\caption{Summary of the fitted parameters for the form factors.}
\resizebox{\linewidth}{!}{
\begin{tabular}{lccccc}
\hline\hline
& $f_1$ & $f_2$ & $f_3$ & $f_1^T$ & $f_2^T$\\
\hline
$f_i(0)$ & $0.54\pm 0.02$& $-0.25\pm 0.02$ & $ 0.74\pm 0.04 $ & $-0.62\pm 0.03$& $-0.35\pm 0.02$\\
$a_1$ & $-0.36\pm 4.44$& $-12.75\pm 0.97$ &$-17.90\pm 7.97$& $-6.88\pm 6.41$& $-7.55\pm 7.62$\\
\hline
& $g_1$ & $g_2$ & $g_3$ & $g_1^T$ & $g_2^T$\\
\hline
$f_i(0)$ & $0.54\pm 0.02$& $-0.25\pm 0.02$ & $ -0.74\pm 0.04 $ & $0.51\pm 0.06$& $-0.35\pm 0.02$\\
$a_1$ & $-3.59\pm 4.45$& $-16.01\pm 0.97$ &$-13.47\pm 8.03$& $-23.25\pm 16.28$& $-10.18\pm 7.84$\\
\hline\hline
\end{tabular}
}
\label{table:fit}
\end{table}

\begin{figure}[th]
\centering
\includegraphics[width=6.5cm]{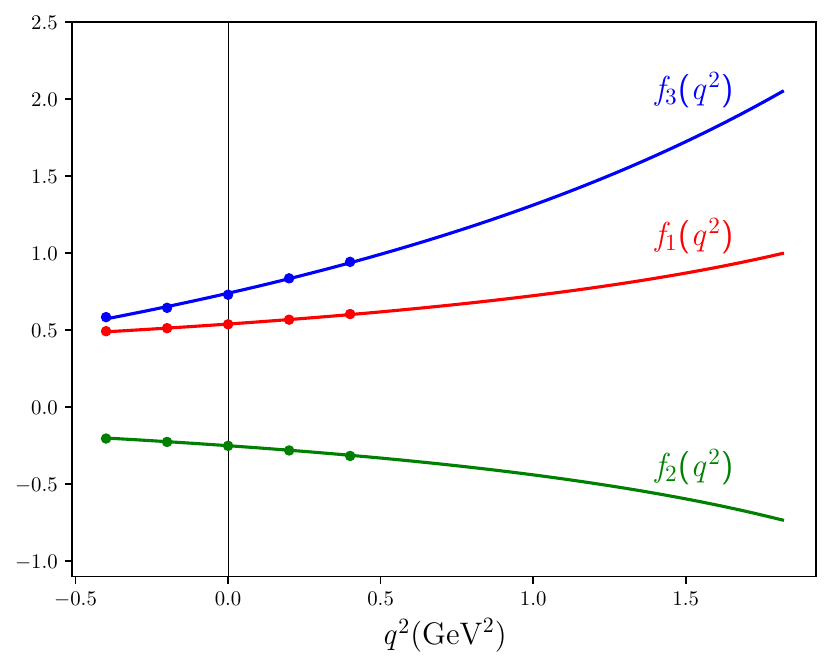}
\includegraphics[width=6.5cm]{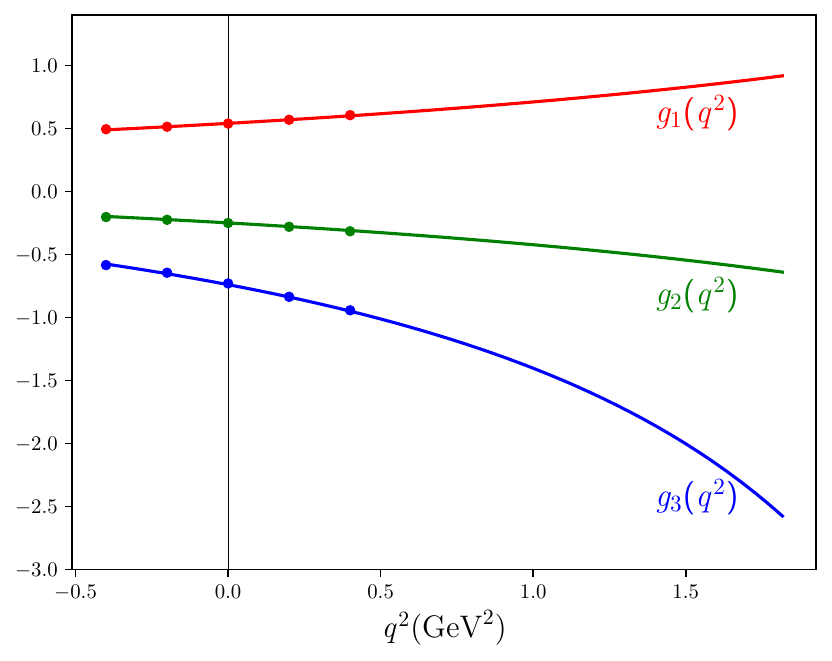}
\includegraphics[width=6.5cm]{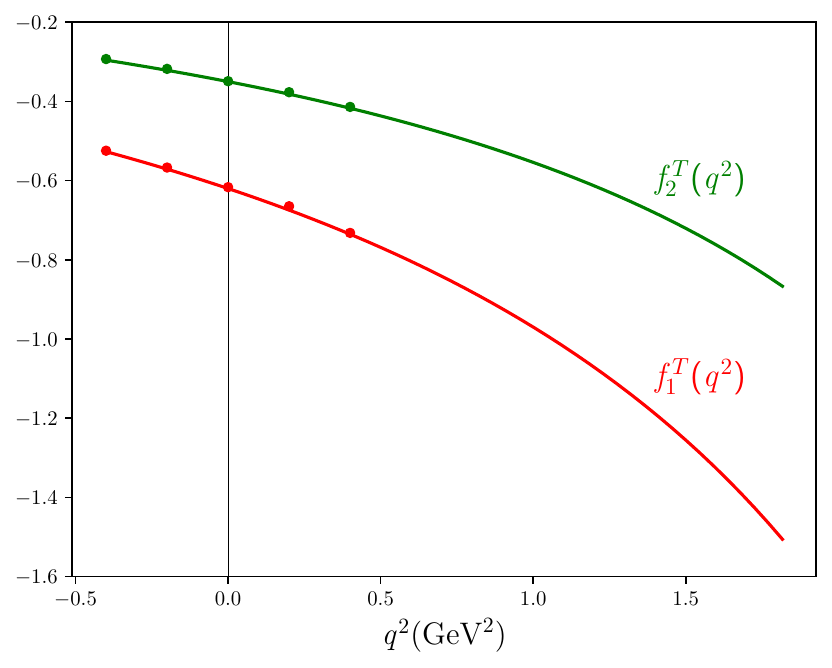}
\includegraphics[width=6.5cm]{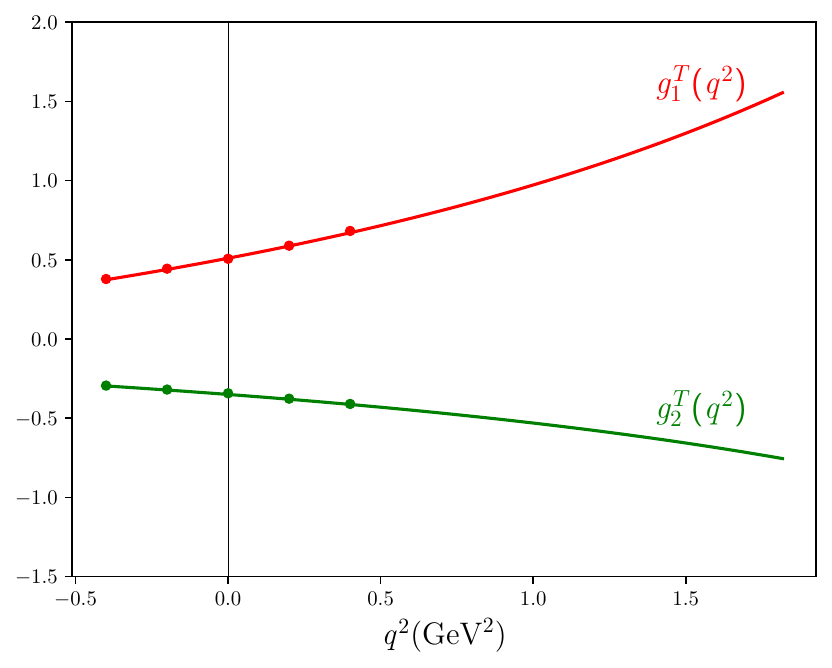}
\caption{The $ q^2 $ dependence of form factors with the central value of fitting parameters listed in Table \ref{table:fit}.  The symbol points denote the fitted points for each of the form factor.}
\label{fig:form-factors}
\end{figure}

To obtain the $ q^2 $ dependence of the form factors, we calculate them over a small spacelike interval $ q^2 \in [-0.4,0.4]$ $\text{GeV}^2  $. Then we fit the data by employing the $ z $-series parametrization~\cite{Bourrely:2008za}, presented as follows:
\begin{align}
\label{BCL}
f_i(q^2)&=\frac{f_i(0)}{1-q^2/(m_{pole})^2}\Bigl\{1+a_1(z(q^2,t_0)-z(0,t_0))\Bigr\}\;,\nonumber\\[10pt]
z(q^2,t_0)&=\frac{\sqrt{t_+-q^2}-\sqrt{t_+-t_0}}{\sqrt{t_+-q^2}+\sqrt{t_+-t_0}}\;,
\end{align}
where $ a_1 $ is a fitting parameter, and $ f_i(0) $ denotes the value of form factors at $ q^2=0 $, which is also treated as a fitted parameter. The variables are defined as $ t_{\pm}= (M_{\Lambda_c}\pm M_p)^2$, and $ t_0=t_+-\sqrt{t_+-t_-} \sqrt{t_+-t_{min}}$. In the numerical analysis, we choose $ t_{min}=-0.4 \,\text{GeV}^2 $. For the pole masses, we use $ M_{pole}=M_{D^*} = 2.01\,\text{GeV}$ for $ f_{1,2}^{(T)} $, $ M_{pole}=M_{D_1} = 2.423\,\text{GeV}$ for $ g_{1,2}^{(T)} $, $ M_{pole}=M_{D_0} = 2.351\,\text{GeV}$ for $ f_3 $, and $ M_{pole}=M_{D} = 1.87\,\text{GeV}$ for $ g_3 $~\cite{ParticleDataGroup:2024cfk}. The nonlinear least squares ($ \chi^2 $) fitting method are employed in our analysis. The fitting results are listed in Table~\ref{table:fit}, where the values of $ f_i(0) $ obtained from the fitting procedure are consistent with our directly calculated results shown in Table~\ref{table:f0}. The $ q^2 $ dependence of the form factors are displayed in Fig.~\ref{fig:form-factors}.

With the obtained form factors, we can now systematically analyzed the branching fractions and corresponding asymmetry observables of the rare decay $ \Lambda_c\to p\ell\ell $.

\subsection{Phenomenological analysis of \texorpdfstring{$\Lambda_c \to p$}{Λc → p} rare decays}

The effective Hamiltonian depicting $ c\to u $ transition can be written as~\cite{Meinel:2017ggx, deBoer:2015boa}
\begin{align}
\mathcal{H}_{e f f}=\frac{G_F\;\alpha}{2\sqrt{2}\pi}&\bigg\{C_9\big[\bar{u} \gamma^\mu\left(1-\gamma_5\right) c\big]\big[\bar{\ell} \gamma_\mu \ell\big]+C_{10}\big[\bar{u} \gamma^\mu\left(1-\gamma_5\right) c\big]\big[\bar{\ell} \gamma_\mu \gamma_5 \ell\big]\nonumber\\[5pt]
&-C_7\frac{2 m_c}{q^2}\big[\bar{u}i\sigma_{\mu\nu}q^\nu(1+\gamma_5)c\big]\big[\bar{\ell} \gamma_\mu \ell\big] \bigg\}\;,
\end{align}
where $ G_F =1.166\times 10^{-5} \;\text{GeV}^{-2}  $ denotes the Fermion constant and $ \alpha= 1/137$ is the fine structure constant. The Wilson coefficient $ C_{10} $ is predicted to be zero within the standard model and $ C_{7,9} $ are functions of $ q^2 $ , which are defined as~\cite{deBoer:2015boa}
\begin{equation}
C_{7,9}(q^2)=\frac{4\pi}{\alpha_s}[V_{cd}^*V_{ud}C_{7,9}^{\mathrm{eff}(d)}(q^2)+V_{cs}^*V_{us}C_{7,9}^{\mathrm{eff}(s)}(q^2)]\;.
\end{equation}
Here, $ V_{q_1q_2} $ are the CKM matrix elements taken from UTFit~\cite{UTfit:2022hsi}. Varying charm scale $m_c / \sqrt{2} \leq \mu_c \leq \sqrt{2} m_c$, the effective coefficient $ C_{7}^{\mathrm{eff}} $ exhibits negligible dependence on $ q^2 $ and falls within the range~\cite{deBoer:2015boa, deBoer:2017que}:
\begin{align}
C_7^{\mathrm{eff}} \in\left[-0.00151-i 0.00556|_s+i 0.00005|_{\text{CKM}},-0.00088-i 0.00327|_s+i 0.00002|_{\text{CKM}}\right]\,,
\end{align}  
where the subscripts denote the contributions to the imaginary part arising from the strong interaction phases and the phases of the CKM matrix.  Another Wilson coefficient $ C_9(q^2) $ is:
\begin{align}
-0.060 X_{d s}(\mu_c=\sqrt{2} m_c) \leq C_9 \leq 0.030 X_{d s}(\mu_c=m_c / \sqrt{2})\,,
\end{align}
where the function $ X_{ds} $ is defined as follows~\cite{deBoer:2015boa}:
\begin{align}
X_{d s}&=\left(V_{c d}^* V_{u d} L(m_d^2, q^2)+V_{c s}^* V_{u s} L(m_s^2, q^2)\right)\,,\nonumber\\[5pt]
L\left(m^2, q^2\right)&=\frac{5}{3}+\ln \frac{\mu_c^2}{m^2}+x-\frac{1}{2}(2+x)|1-x|^{1 / 2} \begin{cases}\ln \frac{1+\sqrt{1-x}}{1-\sqrt{1-x}}-i \pi & x \equiv \frac{(2 m)^2}{q^2}<1 \\ 2 \tan ^{-1}\left[\frac{1}{\sqrt{x-1}}\right] & x \equiv \frac{(2 m)^2}{q^2}>1\end{cases}\,.
\end{align}

The Wilson coefficient $ C_9 $ includes extra perturbation as well as long-distance contributions from the vector resonances $ \rho $, $ \omega $, and $ \phi $ decaying in to lepton pair $ \ell^+\ell^- $, which can be modeled using a usual Breit-Wigner structure:
\begin{equation}
\begin{aligned}C_{9}^{\mathrm{R}}(q^{2})&=a_\omega e^{i\delta_\omega}\left(\frac1{q^2-M_\omega^2+iM_\omega\Gamma_\omega}-\frac3{q^2-M_\rho^2+iM_\rho\Gamma_\rho}\right)+\frac{a_\phi e^{i\delta_\phi}}{q^2-M_\phi^2+iM_\phi\Gamma_\phi}\;,
\end{aligned}
\end{equation}
where the isospin relation is considered to connect the $ \rho $ and $ \omega $ contributions to reduce the theoretical uncertainties~\cite{Golz:2021imq}. $ M_V $ and $ \Gamma_V $ represent the mass and total decay width of the vector mesons $ V = \rho, \omega, \phi $, with the values taken from the Particle Data Group~\cite{ParticleDataGroup:2024cfk}. To fix the two couplings $ a_\omega $ and $ a_\phi $, the following approximation is employed:
\begin{equation}
Br(\Lambda_c(\to p V)\to p\mu^+\mu^-) = Br(\Lambda_c\to p V) Br(V\to \mu^+\mu^-)\;,\,\,\,\,\, V=\omega, \phi\;.
\end{equation}
Here, the left-hand side is directly calculated using only $ C_9^\mathrm{R}(q^2) $, considering only the contributions from the $ \omega $ and $ \phi $, while the right-hand side is evaluated by the latest experimental data~\cite{ParticleDataGroup:2024cfk, LHCb:2024hju}:
\begin{align}
Br(\Lambda_{c}\to p\phi)=(1.06\pm 0.14)\times10^{-3}\;,\,\,\,&Br(\phi\to\mu^{+}\mu^{-})=(2.85\pm 0.22)\times10^{-4}\;,\nonumber\\[5pt]
\frac{\mathcal{B}(\Lambda_c\to p\omega)\mathcal{B}(\omega\to\mu^+\mu^-)}{\mathcal{B}(\Lambda_c\to p\phi)\mathcal{B}(\phi\to\mu^+\mu^-)}&=0.24\pm0.03\pm0.018\;.
\end{align}
As a result, we obtain
\begin{equation}
\begin{aligned}
a_{\omega}&=0.072\pm0.015\;,\,\,\,a_{\phi}&=0.104\pm0.010\;,
\end{aligned}
\end{equation}
which are in excellent agreement with Refs.~\cite{Golz:2021imq, Faustov:2018dkn, Meinel:2017ggx}. Since the strong phases $ \delta_\omega $ and $ \delta_\phi $ are experimentally unknown, we vary them independently in $0$ to $ 2\pi $ range when calculating the decay observables of $ \Lambda_c\to p\ell\ell $. $C_9^\mathrm{R}(q^2)$ can be of order 1, rendering the SM contributions from $C_{7,9}^{\mathrm{eff}}$ negligible across the entire physical region~\cite{Golz:2021imq}. Consequently, we will neglect the effects of $C_{7,9}^{\mathrm{eff}}$ in the analysis of resonance contributions.

\begin{figure}[ht]
\centering
\includegraphics[width=7.9cm]{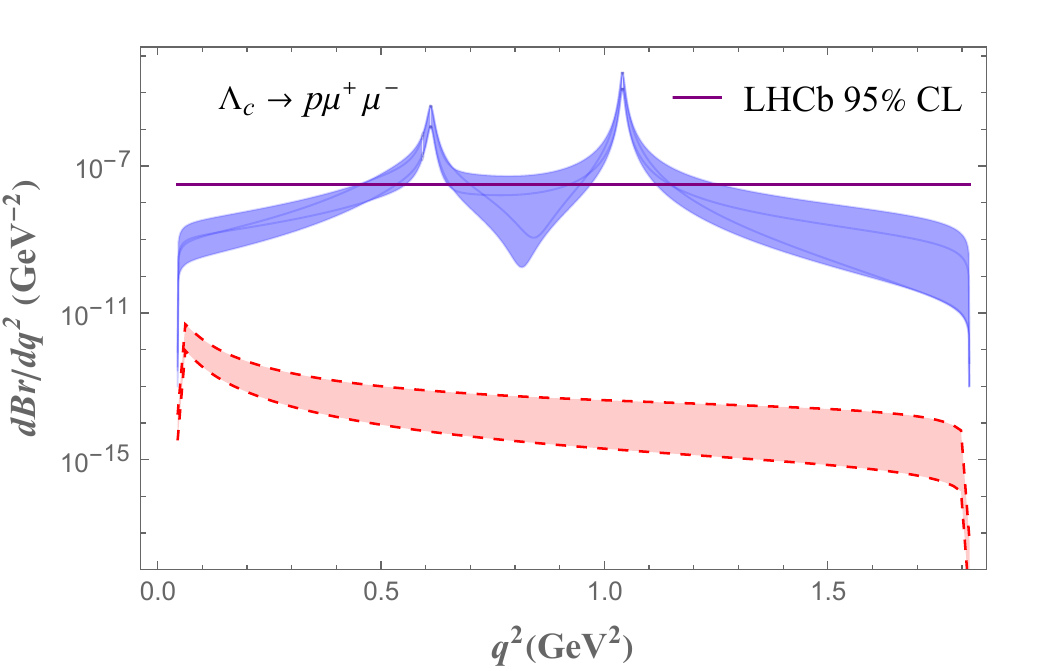}
\includegraphics[width=7.2cm]{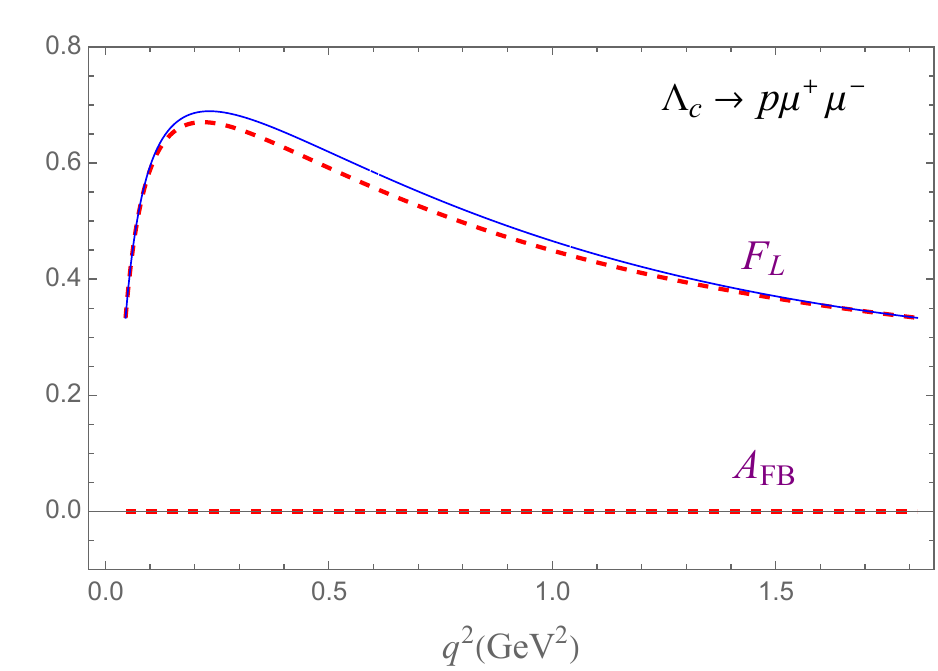}
\caption{Predictions for the $ q^2 $ dependence of $ \mathrm{d}Br/\mathrm{d}q^2 $, $ A_{FB} $ and $ F_L $ for $ \Lambda_c\to p\mu^+\mu^- $ decay mode with and without the vector resonance contributions. The red line represents the contribution excluding resonances, while the red error band reflects uncertainties from the Wilson coefficients and form factors. The blue lines denote the contribution including resonances. The blue error bands represent uncertainties from strong phases $ \delta_\omega $ and $ \delta_\phi $ varying $ 0 $ to $ 2 \pi $, as well as the uncertainties from the couplings $ a_\omega $, $ a_\phi $. The horizontal purple line marks the LHCb upper limit on $ Br(\Lambda_c\to p\mu^+\mu^-) $ in the $ \sqrt{q^2} $ region excluding $ \pm 40\,\mathrm{MeV} $ intervals around $ m_\omega $ and $ m_\phi $.}
\label{fig:observables}
\end{figure}

To further aid in the phenomenological analysis, it is beneficial to introduce helicity amplitudes, as they offer a clearer conceptual framework and streamline the expressions of asymmetric observables. The relations between helicity amplitudes and the form factors are as follows~\cite{Gutsche:2013pp}:
\begin{align}
&H_{\frac{1}{2},0}^{Vm} =\sqrt{\frac{Q_-}{q^2}}\biggl(M_+F_1^{Vm}+\frac{q^2}{M_{\Lambda_c}}F_2^{Vm}\biggr)\;,&& H_{\frac{1}{2},0}^{Am} =\sqrt{\frac{Q_+}{q^2}}\biggl(M_-F_1^{Am}-\frac{q^2}{M_{\Lambda_c}}F_2^{Am}\biggr)\;, \nonumber\\[2pt]
&H_{\frac{1}{2},1}^{Vm} =\sqrt{2Q_-}\biggl(F_1^{Vm}+\frac{M_+}{M_{\Lambda_c}}F_2^{Vm}\biggr)\;, && H_{\frac{1}{2},1}^{Am} =\sqrt{2Q_+}\biggl(F_1^{Am}-\frac{M_-}{M_{\Lambda_c}}F_2^{Am}\biggr)\;, \nonumber\\[2pt]
&H_{\frac{1}{2},t}^{Vm} =\sqrt{\frac{Q_+}{q^2}}\biggl(M_-F_1^{Vm}+\frac{q^2}{M_{\Lambda_c}}F_3^{Vm}\biggr)\;,&&  H_{\frac{1}{2},t}^{Am} =\sqrt{\frac{Q_-}{q^2}}\biggl(M_+F_1^{Am}-\frac{q^2}{M_{\Lambda_c}}F_3^{Am}\biggr)\;.
\end{align}
Here, $ H_{\lambda^\prime, \lambda}^{V(A)m} $ is the helicity amplitudes for weak transitions induced by vector and axial-vector currents, where $ \lambda^\prime $ represents the helicity of the proton and $ \lambda = 0, \pm 1, t $ correspond to longitudinal, transverse, and time-like helicities, respectively. $ Q_{\pm} $ is defined as $ Q_{\pm} = M_{\pm}^2-q^2 $ and $ M_{\pm}=M_{\Lambda_c}\pm M_p $. The form factors $F_i^{Vm}$ and $ F_i^{Am} $ are linear combinations of $ f_i^{(T)} $ and $ g_i^{(T)} $ and also incorporate with the Wilson coefficients, which are defined as:
\begin{align}
&F_{1}^{V1} =C_9^\mathrm{eff}f_1-\frac{2m_c}{M_{\Lambda_c}}C_7^\mathrm{eff}f_1^{T}\;,  & &F_{1}^{A1} =C_9^\mathrm{eff}g_1+\frac{2m_c}{M_{\Lambda_c}}C_7^\mathrm{eff}g_1^{T}\;, \nonumber\\[3pt]
&F_{2}^{V1} =C_9^\mathrm{eff}f_2-\frac{2m_cM_{\Lambda_c}}{q^2}C_7^\mathrm{eff}f_2^{T}\;,  & &F_{2}^{A1} =C_9^\mathrm{eff}g_2+\frac{2m_cM_{\Lambda_c}}{q^2}C_7^\mathrm{eff}g_2^{T}\;, \nonumber\\[3pt]
&F_{3}^{V1} =C_9^\mathrm{eff}f_3+\frac{2m_cM_-}{q^2}C_7^\mathrm{eff}f_1^{T}\;, & &F_{3}^{A1} =C_9^\mathrm{eff}g_3+\frac{2m_cM_+}{q^2}C_7^\mathrm{eff}g_1^{T}\;, \nonumber\\[3pt]
&F_{i}^{V2} =C_{10} f_i\;, & &F_{i}^{A2} =C_{10}g_i\;. 
\end{align}
The negative helicity amplitudes can be derived using the following relations:
\begin{align}
\label{negative-helicity}
H_{-\lambda^\prime,-\lambda}^{Vm}=H_{\lambda^\prime, \lambda}^{Vm}\;, \quad  \quad H_{-\lambda^\prime,-\lambda}^{Am}=-H_{\lambda^\prime, \lambda}^{Am}\;.
\end{align}
Then the total helicity amplitudes can be obtained:
\begin{align}
\label{total-helicity}
H_{\lambda^\prime,\lambda}^m=H_{\lambda^\prime,\lambda}^{Vm}-H_{\lambda^\prime,\lambda}^{Am}\;.
\end{align}
With the above helicity amplitudes, the differential decay width, the lepton forward-backward asymmetry, and the fraction of longitudinally polarized dileptons of $ \Lambda_c\to p\ell\ell $ can be expressed as~\cite{Faustov:2017wbh, Gutsche:2013pp}
\begin{align}
\label{br}
\frac{d\Gamma(\Lambda_c\to p\ell\ell)}{dq^2}&=\frac{\alpha^2\; G_F^2\; q^2\; \sqrt{Q_{+}Q_{-}}}{(2\pi)^5\;48\;M_{\Lambda_c}^3}\sqrt{1-\frac{4m_\ell^2}{q^2}}H_{tot}\;,\\[5pt]
\label{AFB}
A_{FB}^\ell(q^2)&=-\frac34\frac{\sqrt{1-\frac{4m_l^2}{q^2}}H_P^{12}}{H_{tot}}\;,\\[5pt]
\label{FL}
F_L(q^2)&=\frac{\frac12\left(1-\frac{4m_l^2}{q^2}\right)(H_L^{11}+H_L^{22})+\frac{m_l^2}{q^2}(H_U^{11}+H_L^{11}+H_S^{22})}{H_{tot}}\;,
\end{align}
where $ H_{tot} $ is defined as
\begin{align}
\label{Htot}
H_{tot}& =\left.\frac12(H_U^{11}+H_U^{22}+H_L^{11}+H_L^{22})\left(1-\frac{4m_\ell^2}{q^2}\right)+\frac{3m_\ell^2}{q^2}(H_U^{11}+H_L^{11}+H_S^{22})\;,\right. \nonumber\\[5pt]
H_{U}^{mm^{\prime}}& =\left.\text{Re}(H_{\frac{1}{2},1}^mH_{\frac{1}{2},1}^{\dagger m^{\prime}})+\text{Re}(H_{-\frac{1}{2},-1}^mH_{-\frac{1}{2},-1}^{\dagger m^{\prime}})\;,\right. \nonumber\\[5pt]
H_{L}^{mm^{\prime}}& = \text{Re}(H_{\frac{1}{2},0}^mH_{\frac{1}{2},0}^{\dagger m^{\prime}})+\text{Re}(H_{-\frac{1}{2},0}^mH_{-\frac{1}{2},0}^{\dagger m^{\prime}})\;, \nonumber\\[5pt]
H_{S}^{mm^{\prime}}& = \text{Re}(H_{\frac{1}{2},t}^mH_{\frac{1}{2},t}^{\dagger m^{\prime}})+\text{Re}(H_{-\frac{1}{2},t}^mH_{-\frac{1}{2},t}^{\dagger m^{\prime}})\;, \nonumber\\[5pt]
H_{P}^{mm^{\prime}}& = \text{Re}(H_{\frac{1}{2},1}^mH_{\frac{1}{2},1}^{\dagger m'})-\text{Re}(H_{-\frac{1}{2},-1}^mH_{-\frac{1}{2},-1}^{\dagger m'})\;.
\end{align}
The helicity amplitudes show that the lepton forward-backward asymmetry $ A_{FB} $ is proportional to the the Wilson coefficient $ C_{10} $ and thus will vanish within the SM. The form factors $ f_3 $ and $ g_3 $ only appear in the term $ H_S^{22} $, which includes $ C_{10} $, resulting in no contributions to the $ H_{tot} $ within the SM. They are also heavily suppressed by the factor $ m_\ell^2 $. In order to obtain the numerical results of decay observables, the following input parameters are used~\cite{ParticleDataGroup:2024cfk}:
\begin{align}
&m_e=0.511\,\text{MeV}\;,\quad m_\mu=0.106\,\text{GeV}\;,\quad \tau_{\Lambda_c}=(201.5\pm2.7)\times 10^{-15}\,\text{s}\;,
\end{align}
where $ \tau_{\Lambda_c} $ is the mean lifetime of $ \Lambda_c $. By replacing the decay form factors calculated in our model into the helicity expressions, we can derive predictions for the decay observables of $ \Lambda_c\to p\ell\ell $. In Fig.~\ref{fig:observables}, we plot the $ q^2 $ dependence of the differential branching fractions $\mathrm{d}Br/\mathrm{d}q^2\equiv\tau_{\Lambda_c}\mathrm{d}\Gamma/\mathrm{d}q^2 $, $ A_{FB} $, and $ F_L $ for the decay mode $ \Lambda_c\to p\mu^+\mu^- $, with and without the contributions from the long-distance vector resonances $ \rho $, $ \omega $, $ \phi $. The LHCb upper limit of the branching fraction in the $ \sqrt{q^2} $ region, which excludes $ \pm 40 \;\text{MeV} $ intervals around $ m_\omega $ and $ m_\phi $, yields $ Br(\Lambda_c\to p\mu^+\mu^-) < 3.2\times 10^{-8} $ (95\% confidence level)~\cite{LHCb:2024hju} as shown in Fig.~\ref{fig:observables}. We observe that the calculated branching fraction, excluding the resonance contributions, aligns well with the experimental upper limit. The forward-backward asymmetry $ A_{FB} $ is zero across the entire physical region as mentioned earlier. As for the fraction of longitudinally polarized dileptons $ F_L $, we find at both minimal and maximum recoil points $ F_L(q^2=4m^2_\ell) =F_L(q^2=(M_{\Lambda_c}-M_p)^2)=1/3$, which is in agreement with Refs.~\cite{Faustov:2018dkn, Golz:2021imq, Meinel:2017ggx}. This is not a coincidence. At $ q^2 = 4m_\ell^2 $, Eq.~(\ref{FL}) simplifies to $ F_L = \big((H_U^{11}+H_L^{11}+H_S^{22})/4\big)/\big(3(H_U^{11}+H_L^{11}+H_S^{22})/4\big)=1/3 $. At $ q^2 = (M_{\Lambda_c}-M_p)^2 $, $ m_\ell^2 $ becomes negligible, and Eq.~(\ref{FL}) reduces to $ F_L=H_L^{11}/(H_U^{11}+H_L^{11})=1/(1+H_U^{11}/H_L^{11}) $ ($H_S^{22}=H_L^{22}=0$ within the SM). Since $ H_{1/2,0(1)}^{V1} = 0$, we have $ H_U^{11}/H_L^{11} = |H_{1/2,1}^{A1}|^2/|H_{1/2,0}^{A1}|^2 $, following Eqs.~(\ref{negative-helicity}), (\ref{total-helicity}), and (\ref{Htot}). Considering $ q^2 = (M_{\Lambda_c}-M_p)^2 $, $ |H_{1/2,1}^{A1}|^2/|H_{1/2,0}^{A1}|^2 = 2 $, leading to $ F_L =1/3 $. In addition, the global behavior of $ \mathrm{d}Br/\mathrm{d}q^2 $ and $ F_L $ aligns well with the findings in Refs.~\cite{Faustov:2018dkn, Meinel:2017ggx, Golz:2021imq}.

Integrating over the $q^2$ of the curves in Fig.~\ref{fig:observables}, we obtain the total decay width, which allows for determining the decay branching fractions and the mean values of the fraction of longitudinally polarized dileptons $ F_L $. These results are listed in Table~\ref{table:numerical results} and compared with other theoretical predictions. For results including resonance contributions, different theoretical approaches show relatively significant similarity. Our predictions are close to the values obtained by LQCD and RQM, while the results in Ref.\cite{Sirvanli:2016wnr} differ from ours by an order of magnitude. The errors of the branching fractions and $ \langle F_L\rangle $ stem from the uncertainties of the fitted parameters presented in Table~\ref{table:fit}. However, results that exclude resonance contributions exhibit relatively large discrepancies. The observed deviations are not solely attributed to differences in form factors, but are predominantly influenced by variations in the selection of Wilson coefficients. The uncertainties caused by the Wilson coefficients in our results contribute approximately 60\%, indicating that the choice of charm scale $ \mu_c $ plays a critical role in rare decays. The forward-backward asymmetry $A_{FB}$ is predicted to vanish within the SM, thus, any experimental deviation from zero in $A_{FB}$ would signal the presence of new physics. The potential new physics effects will be discussed in Sec.~\ref{sec:NP}.

Other three-body rare charm decays of spin-$1/2$ baryons to a spin-$1/2$ baryon and a lepton pair, such as $ \Xi_c $ decays, can be analyzed in a similar way. This includes processes like $ \Xi_c^0 \to \Sigma^0 \ell \ell $, $ \Xi_c^0 \to \Lambda \ell \ell $, and $ \Xi_c^+ \to \Sigma^+ \ell \ell $. Flavor symmetries are employed to derive the form factors of $ \Xi_c $ rare decays. Specifically, $ \Xi_c^+ \to \Sigma^+ $ is related to $ \Lambda_c \to p $ via U-spin symmetry, which corresponds to the complete exchange of a $d$ quark with an $s$ quark. Due to isospin symmetry, $ \Xi_c^0 \to \Sigma^0 $ includes an additional factor of $ 1/\sqrt{2} $ compared to $ \Xi_c^+ \to \Sigma^+ $. U-spin symmetry then relates $ \Xi_c^0 \to \Sigma^0 $ to $ \Xi_c^0 \to \Lambda $ with a relative factor of $ \sqrt{3} $~\cite{Wang:2021uzi,Dery:2020lbc}. In summary, for any of the ten form factors, denoted here as $ f_{B_i\to B_f} $, we use the relations: $ f_{\Lambda_c \to p} = f_{\Xi_c^+ \to \Sigma^+} = \sqrt{2} f_{\Xi_c^0 \to \Sigma^0} = \sqrt{6} f_{\Xi_c^0 \to \Lambda} $. 

Regarding the couplings $ a_\phi $ and $ a_\omega $, only one relevant two-body decay mode, $ \Xi_c^0 \to \Lambda \phi $, has been observed. The branching fraction for this decay is $ Br(\Xi_c^0 \to \Lambda \phi) = (4.9 \pm 1.5) \times 10^{-4} $~\cite{Belle:2013ntc}, which suggest $ a_\phi=0.20\pm 0.04 $ in $ \Xi_c $ decay mode. There is no sufficient experimental data to determine $ a_\omega $, so we adopt the same value as in the $ \Lambda_c \to p $ process. Applying the same analysis procedure, the branching fractions and the mean values of $ F_L $ of $ \Xi_c $ rare decay mode can be roughly estimated, which are also listed in Table~\ref{table:numerical results}. The input mass and lifetime of $ \Xi_c $ are $ m_{\Xi_c^0} = 2.470\,\text{GeV} $, $ \tau_{\Xi_c^0} = (1.504\pm 0.028)\times 10^{-13}s$, $ m_{\Xi_c^+} = 2.468\,\text{GeV} $, and $ \tau_{\Xi_c^+} = (4.53\pm 0.05)\times 10^{-13}s$~\cite{ParticleDataGroup:2024cfk}.

\begin{table}[ht]
\centering
\caption{Theoretical predictions of the branching fractions, and the mean value of $ F_L $ for the rare decay $ \Lambda_c\to p \ell \ell $ and $ \Xi_c\to (\Sigma,\;\Lambda) \ell \ell $ with and without resonances contributions.}
\resizebox{\linewidth}{!}{
\begin{tabular}{lcccccc}
\hline\hline
& & \multicolumn{2}{c}{$ Br $} & & \multicolumn{2}{c}{$ \langle F_L\rangle $} \\
\hline
& &No-resonance&Resonance & &No-resonance &Resonance \\
\hline
$ \Lambda_c\to p e^+e^- $&  & & &\\
This work &&$ (3.2\pm 2.3)\times 10^{-13} $&$ (4.9\pm 1.4)\times 10^{-7} $  &&$ 0.54\pm 0.01 $& $ 0.55\pm 0.01 $ \\ 
RQM~\cite{Faustov:2018dkn} && $ (3.8\pm 0.5)\times 10^{-12} $&$ (3.7\pm 0.8)\times 10^{-7}$ &&&  \\  
\cite{Sirvanli:2016wnr}& & $(4.5\pm 2.4)\times 10^{-14}$&$ (4.2\pm 0.7)\times 10^{-6} $  && & \\   
\hline
$ \Lambda_c\to p \mu^+\mu^- $& & &   &\\
This work &&$ (2.4\pm 1.8)\times 10^{-13} $&$ (4.9\pm 1.4)\times 10^{-7} $ & &$ 0.47\pm 0.01 $& $ 0.48\pm 0.01 $ \\ 
LQCD~\cite{Meinel:2017ggx}&&$(4.1\pm 0.4^{+6.1}_{-1.9})\times 10^{-11}$&$(3.7\pm 1.3)\times 10^{-7}$ & && \\ 
RQM~\cite{Faustov:2018dkn} && $ (2.8\pm 0.4)\times 10^{-12} $&$(3.7\pm 0.8)\times 10^{-7}$ &&& $ 0.52\pm 0.02 $ \\  
\cite{Sirvanli:2016wnr} & &$(3.77\pm 2.28)\times 10^{-14}$&$ (3.2\pm 0.7)\times 10^{-6} $ & &&  \\ 
\hline 
$ \Xi_c^+\to \Sigma^+ e^+e^- $& & $(6.9\pm 4.9)\times 10^{-13}$&$(3.0\pm 1.5)\times 10^{-6} $ & &$ 0.57\pm 0.02 $&$ 0.57\pm 0.01 $\\
$ \Xi_c^+\to \Sigma^+ \mu^+\mu^- $& & $(5.2\pm 3.7)\times 10^{-13}$&$(3.0\pm 1.5)\times 10^{-6} $ & &$ 0.49\pm 0.02 $&$ 0.49\pm 0.01 $\\
$ \Xi_c^0\to \Sigma^0 e^+e^- $& & $(1.1\pm 0.8)\times 10^{-13}$&$ (4.9\pm 2.4)\times 10^{-7} $ & & $ 0.57\pm 0.02 $&$ 0.57\pm 0.01 $\\
$ \Xi_c^0\to \Sigma^0 \mu^+\mu^- $& &$(8.6\pm 6.2)\times 10^{-14}$&$ (4.9\pm 2.4)\times 10^{-7} $ &  & $ 0.49\pm 0.02 $&$ 0.49\pm 0.01 $ \\
$ \Xi_c^0\to \Lambda e^+e^- $& & $(4.5\pm 2.8)\times 10^{-14}$&$ (2.0\pm 1.0)\times 10^{-7} $ &  &$ 0.56\pm 0.02 $&$ 0.56\pm 0.01 $\\
$ \Xi_c^0\to \Lambda \mu^+\mu^- $& & $(3.4\pm 2.2)\times 10^{-14}$&$ (1.9\pm 1.0)\times 10^{-7} $ & &$ 0.48\pm 0.01 $&$ 0.49\pm 0.01 $\\
\hline\hline
\end{tabular}
}
\label{table:numerical results}
\end{table}

\subsection{Potential new physics effects}\label{sec:NP}

\begin{figure}[ht]
\centering
\includegraphics[width=7cm]{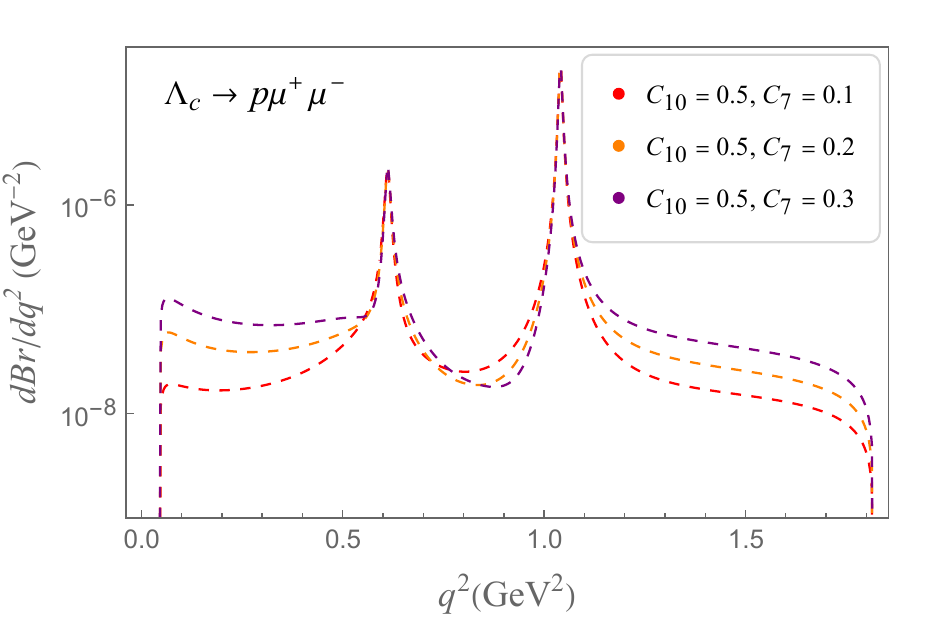}
\includegraphics[width=7cm]{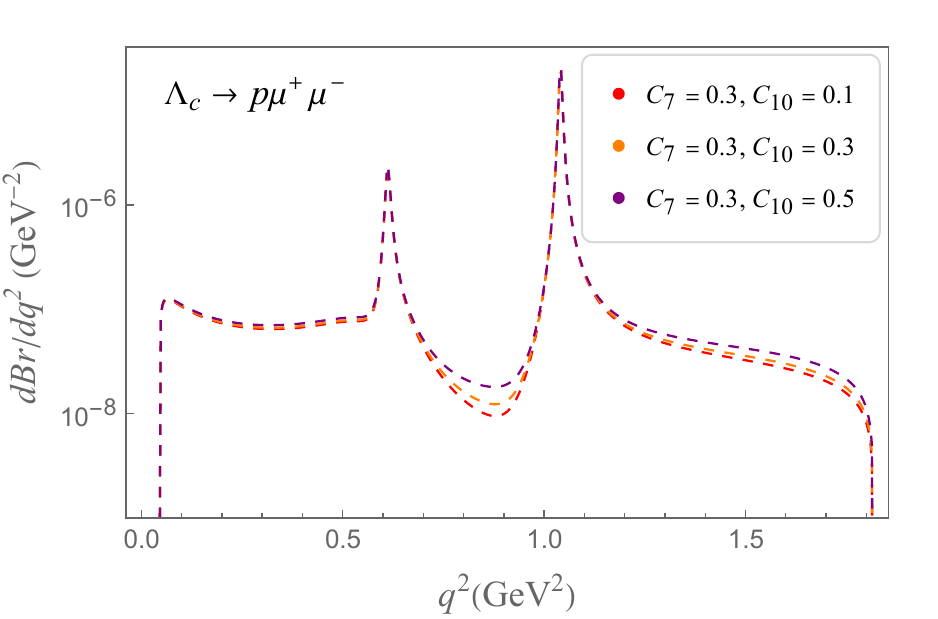}
\includegraphics[width=7cm]{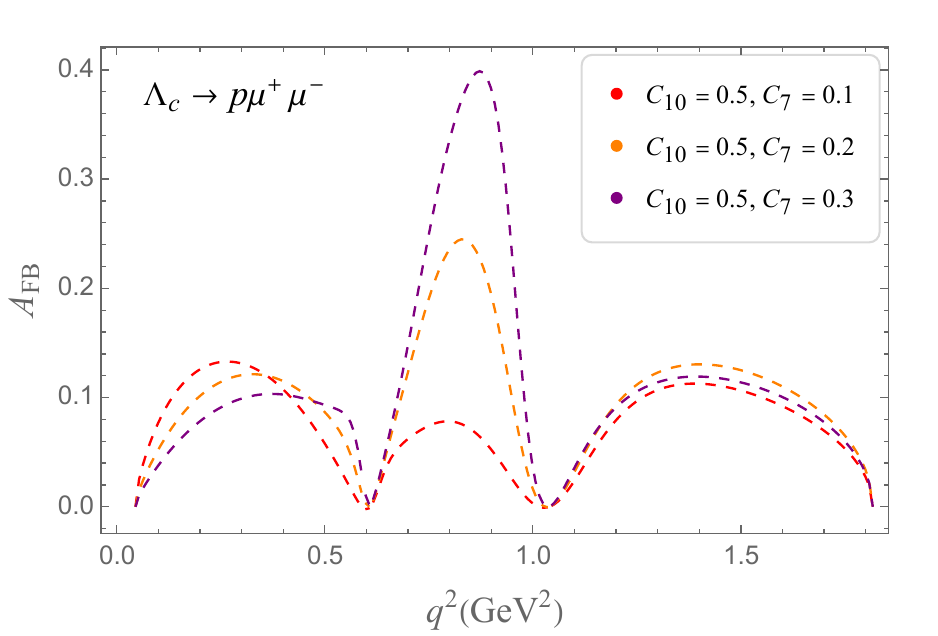}
\includegraphics[width=7cm]{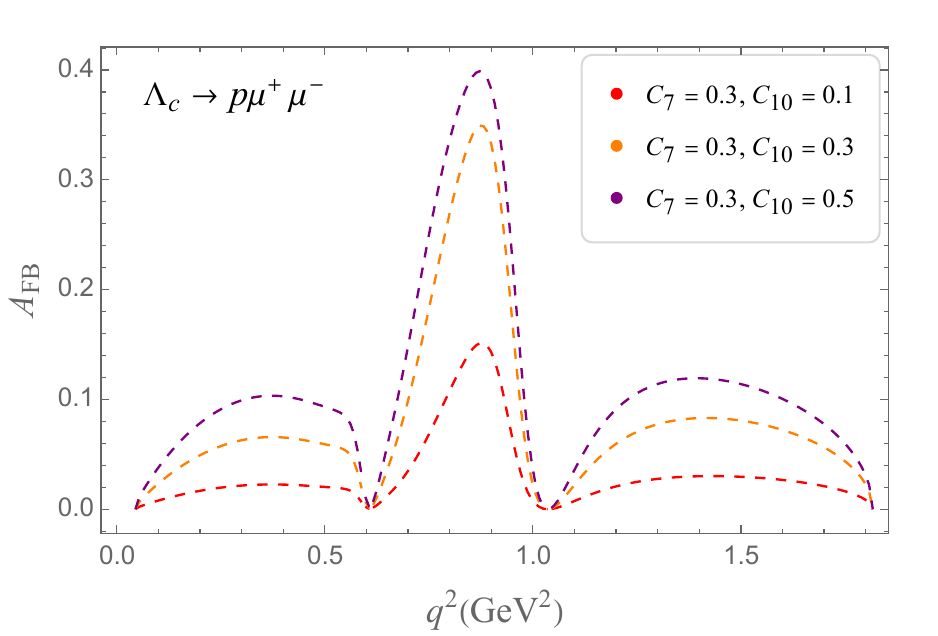}
\includegraphics[width=7cm]{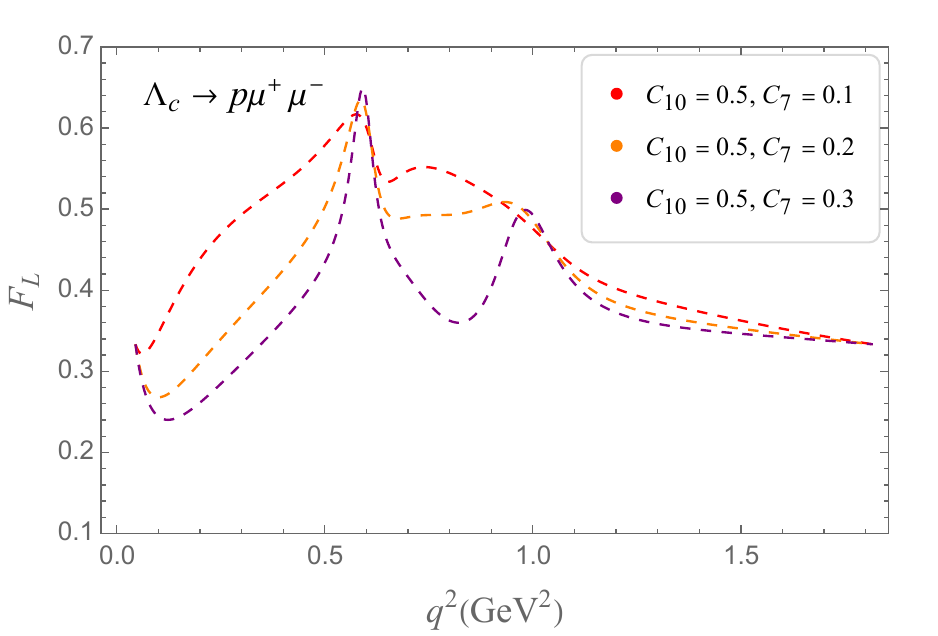}
\includegraphics[width=7cm]{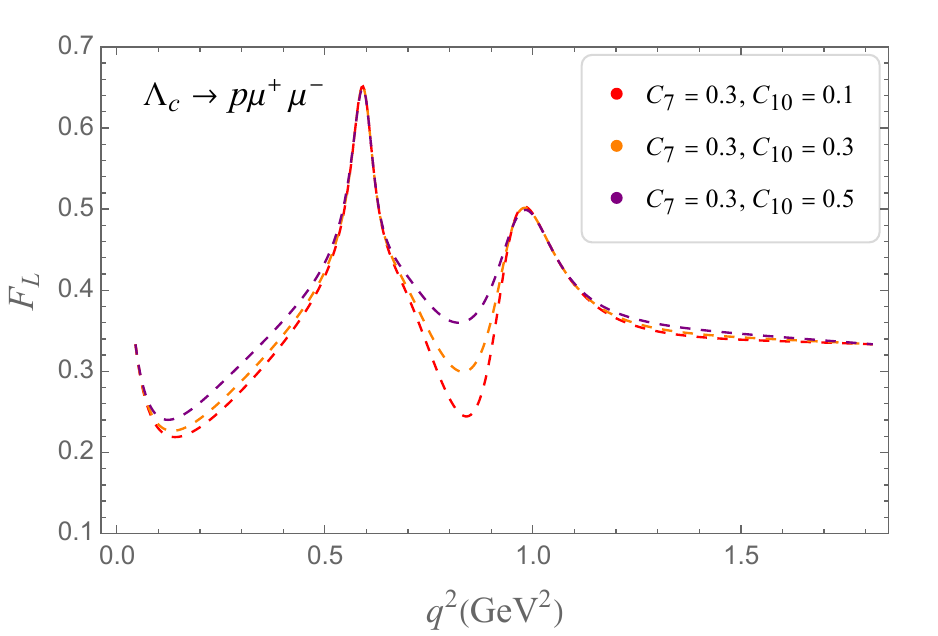}
\caption{New physics effects for the differential branching fraction, the forward-backward asymmetry, and the fraction of longitudinally polarized dileptons with different values of the Wilson coefficients. Strong phases are fixed to $ \delta_\omega = \delta_\phi = 0 $.}
\label{fig:observables-np}
\end{figure}

In this section, we will briefly explore the potential new physics effects through the decay observables in $ \Lambda_c\to p \ell \ell $ decay. For the Wilson coefficients, we use the constraints in Ref.~\cite{Golz:2021imq}, where $ |C_7|\lesssim 0.3 $, $ |C_{10}|\lesssim 0.8 $ and we retain $ C_9 = C_9^R(q^2) $. New physics effects on the differential branching fraction, forward-backward asymmetry, and the fraction of longitudinally polarized dileptons are examined with different Wilson coefficient values, as shown in Fig.~\ref{fig:observables-np}.

The differential branching fraction shows a significant dependence on the parameter $C_7$, especially in the low $q^2$ region. In this region, the branching fraction increases sharply with rising values of $C_7$, indicating a strong correlation between this parameter and the decay process. In contrast, the influence of $C_{10}$ is minimal throughout the entire $q^2$ spectrum, suggesting that variations in $C_{10}$ do not notably affect the branching fractions in the same way as $C_7$ does.

The forward-backward asymmetry $ A_{FB} $ in the $ \Lambda_c\to p \ell \ell $ decay demonstrates notable sensitivity to new physics effects. As illustrated in Fig.~\ref{fig:observables-np}, both $ C_7 $ and $ C_{10} $ substantially affect the behavior of $ A_{FB} $. Notably, $ A_{FB} $ is heavily suppressed near the resonances, where the total helicity amplitude $ H_{tot} $ reaches its peak. Conversely, away from the resonances, even a small value of $C_{10}$, such as 0.1, can lead to appreciable values of $A_{FB}$, making it a clean observable for detecting new physics~\cite{Gisbert:2024kob}. However, the considerable hadronic uncertainties will pose challenges for accurately extracting the Wilson coefficients~\cite{Golz:2021imq}.

The fraction of longitudinally polarized dileptons $ F_L $ exhibits distinct resonance contributions as illustrated Fig.~\ref{fig:observables-np}, which are not apparent within the SM as placed in Fig.~\ref{fig:observables}. The reason for this lack of visibility in the SM is primarily due to the fact that $C_7$ is significantly smaller than $C_9^R(q^2)$. When $C_7$ is effectively neglected, both the numerator and denominator in Eq.~(\ref{FL}) become dominated by the term $|C_9^R|^2$. This results in a cancellation of the contributions from the resonance state, obscuring any potential signals associated with the resonances in $ F_L $. Thus, the observable $F_L$ provides a crucial distinction between potential new physics scenarios and the predictions of the SM.

\section{CONCLUSIONS}\label{Conclusions}

In this work, we investigate the rare decay $\Lambda_c \to p \ell \ell$ within the framework of QCD sum rules. By employing Cutkosky cutting rules and input parameters, we derive the complete expressions of the form factors as well as the numerical values at the maximum recoil point $ q^2 = 0 $. The obtained form factors are extrapolated across the entire physical region using $z$-series parametrizations, which effectively capture the behavior of the $q^2$ dependence. 

Then these form factors are utilized to calculate the branching fractions, the forward-backward asymmetry $A_{FB}$, and the fraction of longitudinally polarized dileptons $F_L$, with and without long-distance contributions from the vector resonances $\rho$, $\omega$, and $\phi$. The resonance contributions are modeled using a Breit-Wigner structure, with coefficients derived from available experimental data. We find that the results including the contributions from resonance are generally similar across different theoretical models, while the results excluding resonance contributions show significant discrepancies. This phenomenon is primarily due to the different choice of Wilson coefficients. Moreover, we provide rough estimates of the branching fractions and $ F_L $ for $ \Xi_c $ rare decays, offering predictions for experimental verification.

Finally, we briefly investigate the potential new physics effects on the differential branching fraction, the forward-backward asymmetry $A_{FB}$, and the fraction of longitudinally polarized dileptons $F_L$. The differential branching fraction shows a strong dependence on $C_7$ in the low $q^2$ region. $A_{FB}$ is zero in the standard model, but becomes non-negligible when $C_{10} \neq 0$, making it a clean test for probing new physics. Additionally, $F_L$ exhibits resonance peaks under new physics effects, absent in the standard model, providing a crucial distinction between new physics and SM, which is detectable in experiment with present data at hand. 

\vspace{0.5cm}
%%%%%%%%%%%%%%%%%%%%%%%%%%%%%%%%%%%%%%%%%%%%%%%%%%%%%%%%%%
{\bf ACKNOWLEDGMENTS}

This work was supported in part by the National Key Research and Development Program of China under Contract No. 2020YFA0406400, by the National Natural Science Foundation of China(NSFC) under the Grants No.~12475087 and No.~12235008.


\begin{thebibliography}{10}

\bibitem{Glashow:1970gm}
S.~L. Glashow, J.~Iliopoulos, and L.~Maiani, {Weak Interactions with Lepton-Hadron Symmetry}, Phys. Rev. D \textbf{2}, 1285--1292 (1970).

\bibitem{Hewett:1996ct}
J.~L. Hewett and J.~D. Wells, {Searching for supersymmetry in rare $B$ decays}, Phys. Rev. D \textbf{55}, 5549--5560 (1997).

\bibitem{Buchalla:2000sk}
G.~Buchalla, G.~Hiller, and G.~Isidori, {Phenomenology of nonstandard $Z$ couplings in exclusive semileptonic $b \to s$ transitions}, Phys. Rev. D \textbf{63}, 014015 (2000).

\bibitem{Bird:2004ts}
C.~Bird, P.~Jackson, R.~V. Kowalewski, and M.~Pospelov, {Search for dark matter in $b \to s$ transitions with missing energy}, Phys. Rev. Lett. \textbf{93}, 201803 (2004).

\bibitem{Davidson:1993qk}
S.~Davidson, D.~C. Bailey, and B.~A. Campbell, {Model independent constraints on leptoquarks from rare processes}, Z. Phys. C \textbf{61}, 613--644 (1994).

\bibitem{Belle:2001oey}
K.~Abe et~al., {Observation of the decay $B \to K \ell^{+} \ell^{-}$}, Phys. Rev. Lett. \textbf{88}, 021801 (2002).

\bibitem{BaBar:2006tnv}
B.~Aubert et~al., {Measurements of branching fractions, rate asymmetries, and angular distributions in the rare decays $B \to K \ell^{+} \ell^{-}$ and $B \to K^{*} \ell^{+} \ell^{-}$}, Phys. Rev. D \textbf{73}, 092001 (2006).

\bibitem{BaBar:2008jdv}
B.~Aubert et~al., {Direct CP, Lepton Flavor and Isospin Asymmetries in the Decays $B \to K^{(*)} \ell^{+} \ell^{-}$}, Phys. Rev. Lett. \textbf{102}, 091803 (2009).

\bibitem{Belle:2009zue}
J.~T. Wei et~al., {Measurement of the Differential Branching Fraction and Forward-Backward Asymmetry for $B \to K^{(*)}\ell^+\ell^-$}, Phys. Rev. Lett. \textbf{103}, 171801 (2009).

\bibitem{LHCb:2012juf}
R.~Aaij et~al., {Differential branching fraction and angular analysis of the $B^{+} \rightarrow K^{+}\mu^{+}\mu^{-}$ decay}, JHEP \textbf{02}, 105 (2013).

\bibitem{LHCb:2014cxe}
R.~Aaij et~al., {Differential branching fractions and isospin asymmetries of $B \to K^{(*)} \mu^+ \mu^-$ decays}, JHEP \textbf{06}, 133 (2014).

\bibitem{BELLE:2019xld}
S.~Choudhury et~al., {Test of lepton flavor universality and search for lepton flavor violation in $B \rightarrow K\ell \ell$ decays}, JHEP \textbf{03}, 105 (2021).

\bibitem{LHCb:2019hip}
R.~Aaij et~al., {Search for lepton-universality violation in $B^+\to K^+\ell^+\ell^-$ decays}, Phys. Rev. Lett. \textbf{122}, 191801 (2019).

\bibitem{BaBar:2011ouc}
J.~P. Lees et~al., {Searches for Rare or Forbidden Semileptonic Charm Decays}, Phys. Rev. D \textbf{84}, 072006 (2011).

\bibitem{LHCb:2013hxr}
R.~Aaij et~al., {Search for $D^+_{s} \to \pi^+ \mu^+ \mu^-$ and $D^+_{s} \to \pi^- \mu^+ \mu^+$ decays}, Phys. Lett. B \textbf{724}, 203--212 (2013).

\bibitem{LHCb:2020car}
R.~Aaij et~al., {Searches for 25 rare and forbidden decays of $D^{+}$ and $ {D}_s^{+} $ mesons}, JHEP \textbf{06}, 044 (2021).

\bibitem{CLEO:2010ksb}
P.~Rubin et~al., {Search for rare and forbidden decays of charm and charmed-strange mesons to final states $h^+- e^-+ e^+$}, Phys. Rev. D \textbf{82}, 092007 (2010).

\bibitem{D0:2007qbl}
V.~M. Abazov et~al., {Search for flavor-changing-neutral-current $D$ meson decays}, Phys. Rev. Lett. \textbf{100}, 101801 (2008).

\bibitem{NA482:2009pfe}
J.~R. Batley et~al., {Precise measurement of the $ K^{\pm} \to \pi^{\pm}e^+e^- $ decay}, Phys. Lett. B \textbf{677}, 246--254 (2009).

\bibitem{NA62:2022qes}
E.~Cortina~Gil et~al., {A measurement of the $K^{+} \to \pi^{+} \mu^{+} \mu^{-}$ decay}, JHEP \textbf{11}, 011 (2022), [Addendum: JHEP 06, 040 (2023)].

\bibitem{Mannel:1997xy}
T.~Mannel and S.~Recksiegel, {Flavor changing neutral current decays of heavy baryons: The Case $\Lambda_b \to \Lambda \gamma$}, J. Phys. G \textbf{24}, 979--990 (1998).

\bibitem{Wang:2021uzi}
R.-M. Wang, Y.-G. Xu, C.~Hua, and X.-D. Cheng, {Studying $\mathcal{B}_1(\frac{1}{2}^+)\to \mathcal{B}_2(\frac{1}{2}^+)\ell^+\ell^-$ semileptonic weak baryon decays with the SU(3) flavor symmetry}, Phys. Rev. D \textbf{103}, 013007 (2021).

\bibitem{CDF:2011buy}
T.~Aaltonen et~al., {Observation of the Baryonic Flavor-Changing Neutral Current Decay $\Lambda_{b} \to \Lambda \mu^{+} \mu^{-}$}, Phys. Rev. Lett. \textbf{107}, 201802 (2011).

\bibitem{LHCb:2013uqx}
R.~Aaij et~al., {Measurement of the differential branching fraction of the decay $\Lambda_b^0\rightarrow\Lambda\mu^+\mu^-$}, Phys. Lett. B \textbf{725}, 25--35 (2013).

\bibitem{LHCb:2017rdd}
R.~Aaij et~al., {Evidence for the rare decay $\Sigma^+ \to p \mu^+ \mu^-$}, Phys. Rev. Lett. \textbf{120}, 221803 (2018).

\bibitem{NA48:2007smd}
J.~R. Batley et~al., {First observation and branching fraction and decay parameter measurements of the weak radiative decay $ \Xi^0\to \Lambda e^+e^- $}, Phys. Lett. B \textbf{650}, 1--8 (2007).

\bibitem{LHCb:2024hju}
R.~Aaij et~al., {Search for the rare decay of charmed baryon $ \Lambda_c $ into the $ p\mu^+\mu^- $ final state}, Phys. Rev. D \textbf{110}, 052007 (2024).

\bibitem{LHCb:2017yqf}
R.~Aaij et~al., {Search for the rare decay $\Lambda_{c}^{+} \to p\mu^+\mu^-$}, Phys. Rev. D \textbf{97}, 091101 (2018).

\bibitem{deBoer:2015boa}
S.~de~Boer and G.~Hiller, {Flavor and new physics opportunities with rare charm decays into leptons}, Phys. Rev. D \textbf{93}, 074001 (2016).

\bibitem{Fajfer:2001sa}
S.~Fajfer, S.~Prelovsek, and P.~Singer, {Rare charm meson decays $D \to P l^+ l^-$ and $c \to u l^+ l^-$ in SM and MSSM}, Phys. Rev. D \textbf{64}, 114009 (2001).

\bibitem{Paul:2011ar}
A.~Paul, I.~I. Bigi, and S.~Recksiegel, {On $D\to X_u l^+ l^-$ within the Standard Model and Frameworks like the Littlest Higgs Model with T Parity}, Phys. Rev. D \textbf{83}, 114006 (2011).

\bibitem{Golz:2021imq}
M.~Golz, G.~Hiller, and T.~Magorsch, {Probing for new physics with rare charm baryon (\ensuremath{\Lambda}$_{c}$, \ensuremath{\Xi}$_{c}$, \ensuremath{\Omega}$_{c}$) decays}, JHEP \textbf{09}, 208 (2021).

\bibitem{LHCb:2020gog}
R.~Aaij et~al., {Angular Analysis of the $B^{+}\rightarrow K^{\ast+}\mu^{+}\mu^{-}$ Decay}, Phys. Rev. Lett. \textbf{126}, 161802 (2021).

\bibitem{Wang:2008sm}
Y.-M. Wang, Y.~Li, and C.-D. Lu, {Rare Decays of $\Lambda_b \to \Lambda + \gamma$ and $\Lambda_b \to \Lambda + l^+ l^-$ in the Light-cone Sum Rules}, Eur. Phys. J. C \textbf{59}, 861--882 (2009).

\bibitem{Cheng:1994kp}
H.-Y. Cheng, C.-Y. Cheung, G.-L. Lin, Y.~C. Lin, T.-M. Yan, and H.-L. Yu, {Effective Lagrangian approach to weak radiative decays of heavy hadrons}, Phys. Rev. D \textbf{51}, 1199--1214 (1995).

\bibitem{Faustov:2017wbh}
R.~N. Faustov and V.~O. Galkin, {Rare $\Lambda_b\to\Lambda l^+l^-$ and $\Lambda_b\to\Lambda\gamma$ decays in the relativistic quark model}, Phys. Rev. D \textbf{96}, 053006 (2017).

\bibitem{Faustov:2017ous}
R.~N. Faustov and V.~O. Galkin, {Rare $\Lambda_b\to n l^+l^-$ decays in the relativistic quark-diquark picture}, Mod. Phys. Lett. A \textbf{32}, 1750125 (2017).

\bibitem{Faustov:2018dkn}
R.~N. Faustov and V.~O. Galkin, {Rare $\Lambda _c\rightarrow p \ell ^+\ell ^-$ decay in the relativistic quark model}, Eur. Phys. J. C \textbf{78}, 527 (2018).

\bibitem{Gutsche:2013pp}
T.~Gutsche, M.~A. Ivanov, J.~G. Korner, V.~E. Lyubovitskij, and P.~Santorelli, {Rare baryon decays $\Lambda_b \to \Lambda {l^{+}l^{-}} (l=e, \mu, \tau)$ and $\Lambda_b \to \Lambda\gamma$ : differential and total rates, lepton- and hadron-side forward-backward asymmetries}, Phys. Rev. D \textbf{87}, 074031 (2013).

\bibitem{Liu:2019igt}
L.-L. Liu, X.-W. Kang, Z.-Y. Wang, and X.-H. Guo, {Rare $\Lambda_b \rightarrow \Lambda \ell^+ \ell^- $ decay in the Bethe-Salpeter equation approach}, Chin. Phys. C \textbf{44}, 083107 (2020).

\bibitem{Liu:2019rpm}
L.-L. Liu, C.~Wang, X.-W. Kang, and X.-H. Guo, {FCNC transitions of $\Lambda_b$ to neutron in Bethe-Salpeter equation approach}, Eur. Phys. J. C \textbf{80}, 193 (2020).

\bibitem{Azizi:2010zzb}
K.~Azizi, M.~Bayar, Y.~Sarac, and H.~Sundu, {FCNC transitions of $\Lambda_{b,c}$ to nucleon in SM}, J. Phys. G \textbf{37}, 115007 (2010).

\bibitem{Gan:2012tt}
L.-F. Gan, Y.-L. Liu, W.-B. Chen, and M.-Q. Huang, {Improved Light-cone QCD Sum Rule Analysis Of The Rare Decays $\Lambda_b\rightarrow\Lambda\gamma$ And $\Lambda_b\rightarrow\Lambda l^+l^-$}, Commun. Theor. Phys. \textbf{58}, 872--882 (2012).

\bibitem{Aliev:2018hyy}
T.~M. Aliev, T.~Barakat, and M.~Savc\i{}, {Form factors for the rare $ \Lambda_b(\Lambda_b^*) \to N\ell^+\ell^-$ decays in light cone QCD sum rules}, Phys. Rev. D \textbf{98}, 035033 (2018).

\bibitem{Detmold:2016pkz}
W.~Detmold and S.~Meinel, {$\Lambda_b \to \Lambda \ell^+ \ell^-$ form factors, differential branching fraction, and angular observables from lattice QCD with relativistic $b$ quarks}, Phys. Rev. D \textbf{93}, 074501 (2016).

\bibitem{Meinel:2017ggx}
S.~Meinel, {$\Lambda_c \to N$ form factors from lattice QCD and phenomenology of $\Lambda_c \to n \ell^+ \nu_\ell$ and $\Lambda_c \to p \mu^+ \mu^-$ decays}, Phys. Rev. D \textbf{97}, 034511 (2018).

\bibitem{Hussain:1992rb}
F.~Hussain, D.-S. Liu, M.~Kramer, J.~G. Korner, and S.~Tawfiq, {General analysis of weak decay form-factors in heavy to heavy and heavy to light baryon transitions}, Nucl. Phys. B \textbf{370}, 259--277 (1992).

\bibitem{Mannel:1990vg}
T.~Mannel, W.~Roberts, and Z.~Ryzak, {Baryons in the heavy quark effective theory}, Nucl. Phys. B \textbf{355}, 38--53 (1991).

\bibitem{Chen:2001ki}
C.-H. Chen and C.~Q. Geng, {Rare $\Lambda_b \to \Lambda l^+ l^-$ decays with polarized lambda}, Phys. Rev. D \textbf{63}, 114024 (2001).

\bibitem{Huang:1998ek}
C.-S. Huang and H.-G. Yan, {Exclusive rare decays of heavy baryons to light baryons: $ \Lambda_b \rightarrow \Lambda \gamma $ and $ \Lambda_b \rightarrow \Lambda l^{+} l^{-} $}, Phys. Rev. D \textbf{59}, 114022 (1999), [Erratum: Phys.Rev.D 61, 039901 (2000)].

\bibitem{Lian:2023cgs}
D.-K. Lian, W.~Chen, H.-X. Chen, L.-Y. Dai, and T.~G. Steele, {Strong decays of $T^a_{c{\bar{s}0}}(2900)^{++/0}$ as a fully open-flavor tetraquark state}, Eur. Phys. J. C \textbf{84}, 1 (2024).

\bibitem{Wang:2016wkj}
Z.-G. Wang, {Analysis of the strong decay $X(5568) \rightarrow B_s^0\pi ^+$ with QCD sum rules}, Eur. Phys. J. C \textbf{76}, 279 (2016).

\bibitem{Dias:2013qga}
J.~M. Dias, X.~Liu, and M.~Nielsen, {Predicition for the decay width of a charged state near the $D_s\bar{D}^*/D^*_s\bar{D}$ threshold}, Phys. Rev. D \textbf{88}, 096014 (2013).

\bibitem{Wan:2020oxt}
B.-D. Wan and C.-F. Qiao, {About the exotic structure of $Z_{cs}$}, Nucl. Phys. B \textbf{968}, 115450 (2021).

\bibitem{Wang:2023sii}
Z.-G. Wang, {Decipher the width of the X(3872) via the QCD sum rules}, Phys. Rev. D \textbf{109}, 014017 (2024).

\bibitem{Dai:1996xv}
Y.-B. Dai, C.-S. Huang, M.-Q. Huang, and C.~Liu, {QCD sum rule analysis for the $ \Lambda_b $ $ \rightarrow $ $ \Lambda_c $ semileptonic decay}, Phys. Lett. B \textbf{387}, 379--385 (1996).

\bibitem{Zhang:2023nxl}
S.-Q. Zhang and C.-F. Qiao, {$\Lambda_c$ semileptonic decays}, Phys. Rev. D \textbf{108}, 074017 (2023).

\bibitem{Huang:1998rq}
C.-S. Huang, C.-F. Qiao, and H.-G. Yan, {Decay $ \Lambda_b $ $ \rightarrow $ $p l \bar{\nu}$ in QCD sum rules}, Phys. Lett. B \textbf{437}, 403--407 (1998).

\bibitem{MarquesdeCarvalho:1999bqs}
R.~S. Marques~de Carvalho, F.~S. Navarra, M.~Nielsen, E.~Ferreira, and H.~G. Dosch, {Form-factors and decay rates for heavy Lambda semileptonic decays from QCD sum rules}, Phys. Rev. D \textbf{60}, 034009 (1999).

\bibitem{Shi:2019hbf}
Y.-J. Shi, W.~Wang, and Z.-X. Zhao, {QCD Sum Rules Analysis of Weak Decays of Doubly-Heavy Baryons}, Eur. Phys. J. C \textbf{80}, 568 (2020).

\bibitem{Zhao:2020mod}
Z.-X. Zhao, R.-H. Li, Y.-L. Shen, Y.-J. Shi, and Y.-S. Yang, {The semi-leptonic form factors of $\Lambda_{b}\to\Lambda_{c}$ and $\Xi_{b}\to\Xi_{c}$ in QCD sum rules}, Eur. Phys. J. C \textbf{80}, 1181 (2020).

\bibitem{Zhao:2021sje}
Z.-X. Zhao, X.-Y. Sun, F.-W. Zhang, Y.-P. Xing, and Y.-T. Yang, {Semileptonic form factors of $ \Xi_c\to \Xi $ in QCD sum rules}, Phys. Rev. D \textbf{108}, 116008 (2023).

\bibitem{Xing:2021enr}
Z.-P. Xing and Z.-X. Zhao, {QCD sum rules analysis of weak decays of doubly heavy baryons: the $b\rightarrow c$ processes}, Eur. Phys. J. C \textbf{81}, 1111 (2021).

\bibitem{zhang:2024ick}
S.-Q. Zhang, X.-H. Zhang, and C.-F. Qiao, {Hyperon semileptonic decays in QCD sum rules}, JHEP \textbf{06}, 122 (2024).

\bibitem{Xing:2018lre}
Z.-P. Xing and Z.-X. Zhao, {Weak decays of doubly heavy baryons: the FCNC processes}, Phys. Rev. D \textbf{98}, 056002 (2018).

\bibitem{Chung:1981wm}
Y.~Chung, H.~G. Dosch, M.~Kremer, and D.~Schall, {{QCD} Sum Rules for 'Baryonic Currents'}, Phys. Lett. B \textbf{102}, 175--179 (1981).

\bibitem{Wang:2012hu}
Z.-G. Wang, {Semileptonic decays $B_c^* \to \eta_c \ell \bar{\nu}_{\ell} $ with QCD sum rules}, Commun. Theor. Phys. \textbf{61}, 81--88 (2014).

\bibitem{Yang:2005bv}
M.-Z. Yang, {Semileptonic decay of $B$ and $D \to K_0^*(1430)\bar{\ell} \nu$ from QCD sum rule}, Phys. Rev. D \textbf{73}, 034027 (2006), [Erratum: Phys.Rev.D 73, 079901 (2006)].

\bibitem{Du:2003ja}
D.-S. Du, J.-W. Li, and M.-Z. Yang, {Form-factors and semileptonic decay of $D^+_s \to\phi\bar{l}\nu$ from QCD sum rule}, Eur. Phys. J. C \textbf{37}, 173--184 (2004).

\bibitem{Shifman:1978by}
M.~A. Shifman, A.~I. Vainshtein, and V.~I. Zakharov, {QCD and Resonance Physics: Applications}, Nucl. Phys. B \textbf{147}, 448--518 (1979).

\bibitem{Wan:2022xkx}
B.-D. Wan, S.-Q. Zhang, and C.-F. Qiao, {Possible structure of the newly found exotic state $\eta_1(1855)$}, Phys. Rev. D \textbf{106}, 074003 (2022).

\bibitem{Colangelo:2000dp}
P.~Colangelo and A.~Khodjamirian, {QCD sum rules, a modern perspective}, At The Frontier of Particle Physics 1495--1576 (2000).

\bibitem{ParticleDataGroup:2024cfk}
S.~Navas et~al., {Review of particle physics}, Phys. Rev. D \textbf{110}, 030001 (2024).

\bibitem{Wan:2021vny}
B.-D. Wan, S.-Q. Zhang, and C.-F. Qiao, {Light baryonium spectrum}, Phys. Rev. D \textbf{105}, 014016 (2022).

\bibitem{Khodjamirian:2011jp}
A.~Khodjamirian, C.~Klein, T.~Mannel, and Y.~M. Wang, {Form Factors and Strong Couplings of Heavy Baryons from QCD Light-Cone Sum Rules}, JHEP \textbf{09}, 106 (2011).

\bibitem{Chung:1984gr}
Y.~Chung, H.~G. Dosch, M.~Kremer, and D.~Schall, {Chiral Symmetry Breaking Condensates for Baryonic Sum Rules}, Z. Phys. C \textbf{25}, 151 (1984).

\bibitem{Wan:2024fam}
B.-D. Wan and H.-T. Xu, {0$^{--}$ hidden-heavy tetraquark states via QCD sum rules}, Chin. Phys. C \textbf{48}, 093103 (2024).

\bibitem{Wan:2023epq}
B.-D. Wan, {Mass spectra of $0^{--}$ and $0^{+-}$ hidden-heavy baryoniums}, Eur. Phys. J. C \textbf{84}, 760 (2024).

\bibitem{Zhang:2022obn}
S.-Q. Zhang, B.-D. Wan, L.~Tang, and C.-F. Qiao, {Gluonic nature of the newly observed state X(2600)}, Phys. Rev. D \textbf{106}, 074010 (2022).

\bibitem{Leljak:2019fqa}
D.~Leljak and B.~Melic, {$|V_{ub}|$ determination and testing of lepton flavour universality in semileptonic $B_{c} \to D^{\ast }$ decays}, JHEP \textbf{02}, 171 (2020).

\bibitem{Ball:1997rj}
P.~Ball and V.~M. Braun, {Use and misuse of QCD sum rules in heavy to light transitions: The Decay $ B\to \rho e\nu $ reexamined}, Phys. Rev. D \textbf{55}, 5561--5576 (1997).

\bibitem{Khodjamirian:1997lay}
A.~Khodjamirian and R.~Ruckl, {QCD sum rules for exclusive decays of heavy mesons}, Adv. Ser. Direct. High Energy Phys. \textbf{15}, 345--401 (1998).

\bibitem{Bourrely:2008za}
C.~Bourrely, I.~Caprini, and L.~Lellouch, {Model-independent description of $ B\rightarrow \pi l \nu $ decays and a determination of $|V_{ub}|$}, Phys. Rev. D \textbf{79}, 013008 (2009), [Erratum: Phys.Rev.D 82, 099902 (2010)].

\bibitem{UTfit:2022hsi}
M.~Bona et~al., {New UTfit Analysis of the Unitarity Triangle in the Cabibbo-Kobayashi-Maskawa scheme}, Rend. Lincei Sci. Fis. Nat. \textbf{34}, 37--57 (2023).

\bibitem{deBoer:2017que}
S.~de~Boer and G.~Hiller, {Rare radiative charm decays within the standard model and beyond}, JHEP \textbf{08}, 091 (2017).

\bibitem{Sirvanli:2016wnr}
B.~B. \c{S}irvanli, {Search for $c \to ul^+l^-$ transition in charmed baryon decays}, Phys. Rev. D \textbf{93}, 034027 (2016).

\bibitem{Dery:2020lbc}
A.~Dery, M.~Ghosh, Y.~Grossman, and S.~Schacht, {SU(3)$_{F}$ analysis for beauty baryon decays}, JHEP \textbf{03}, 165 (2020).

\bibitem{Belle:2013ntc}
R.~Chistov et~al., {First observation of Cabibbo-suppressed $\Xi_c^0$ decays}, Phys. Rev. D \textbf{88}, 071103 (2013).

\bibitem{Gisbert:2024kob}
H.~Gisbert, G.~Hiller, and D.~Suelmann, {Effective field theory analysis of rare $|\Delta c| = |\Delta u| = 1$ charm decays}, JHEP \textbf{12}, 102 (2024).

\end{thebibliography}
\end{document}